\documentclass[10pt,twocolumn,reqno, aps,prx,superscriptaddress]{revtex4-1}%
\usepackage{amsmath}
\usepackage[pdftex]{graphicx}
\usepackage{float}
\usepackage{color}
\usepackage{rotating}
\usepackage[breaklinks=true,colorlinks,citecolor=blue,linkcolor=blue,urlcolor=blue]%
{hyperref}
\usepackage{graphicx}
\usepackage{natbib}
\usepackage{dcolumn}
\usepackage{bm}
\usepackage{hyperref}
\usepackage[caption=false]{subfig}
\usepackage{epstopdf}
\usepackage{ulem}
\usepackage{amsfonts}
\usepackage{amssymb}%
\providecommand{\U}[1]{\protect\rule{.1in}{.1in}}
\providecommand{\U}[1]{\protect\rule{.1in}{.1in}}
\begin{document}

\title{Resources of nonlinear cavity magnonics for quantum information}
\author{Mehrdad Elyasi}
%\email{melyasi@imr.tohoku.ac.jp}
\affiliation{Institute for Materials Research, Tohoku University, 2-1-1 Katahira, 980-8577 Sendai, Japan
}
\author{Yaroslav M. Blanter}
%\email{ymblanter@tudelft.nl}
\affiliation{Kavli Institute of Nanoscience, Delft University of Technology,
Lorentzweg 1, 2628CJ Delft, The Netherlands}
\author{Gerrit E. W. Bauer}
\affiliation{Zernike Institute for Advanced Materials, University of Groningen, The Netherlands}
\affiliation{Institute for Materials Research $\text{\&}$ AIMR \& CSRN, Tohoku University, 980-8577 Sendai, Japan}

\begin{abstract}
We theoretically explore nonlinearities of ferromagnets in microwave cavities
in the classical and quantum regimes, and assess the resources for quantum
information, i.e. fluctuation squeezing and bipartite entanglement. The
(semi-)classical analysis of the anharmonic oscillator (Duffing) model for the
Kittel mode when including all other magnon modes, reveals chaotic and limit-cycle phases that do not survive in
quantum calculations. However, magnons with nonzero wavenumbers that are
driven by the Suhl instability of the Kittel mode, form a genuine limit cycle.
We subsequently compute bounds for the distillable entanglement, as well as entanglement of
formation for the bipartite configurations of the mixed magnon modes. The
distillable entanglement of bipartite states accessible from a covariance
matrix vanishes, but can be recovered by injection locking. The predicted
magnon entanglement can be experimentally tested with yttrium iron garnet
samples under realistic conditions.

\end{abstract}
\maketitle

\preprint{APS/123-QED}

\section{Introduction}

Cavity optomagnonics is the emergent field devoted to understand the
interaction of magnons --- the quanta of the elementary spin wave excitations
of the magnetic order --- with electromagnetic waves confined to cavities
\cite{Zhang2014,Tabuchi2014}. While optomagnonic coupling to (infrared) light
is dispersive and, at least to date, rather weak
\cite{Osada2016,Zhang2016,Haigh2016,Sharma2017}, magnons (ultra) strongly
interact with microwave (MW) photons \cite{Huebl2013,Zhang2014,Tabuchi2014},
thereby enabling classical and quantum information processing and storage with
coherently controlled magnons
\cite{Raimond2001,Mabuchi2002,Xiang2013,Tabuchi2015}. Up/down-quantum
converters between both communication (optical fibers) and processing
(superconducting qubits) units have been envisioned and pursued
\cite{Haigh2015,Zhang2015,Bai2015,Osada2016,Zhang2016}.

\begin{figure}[!]
\includegraphics[width=0.5\textwidth]{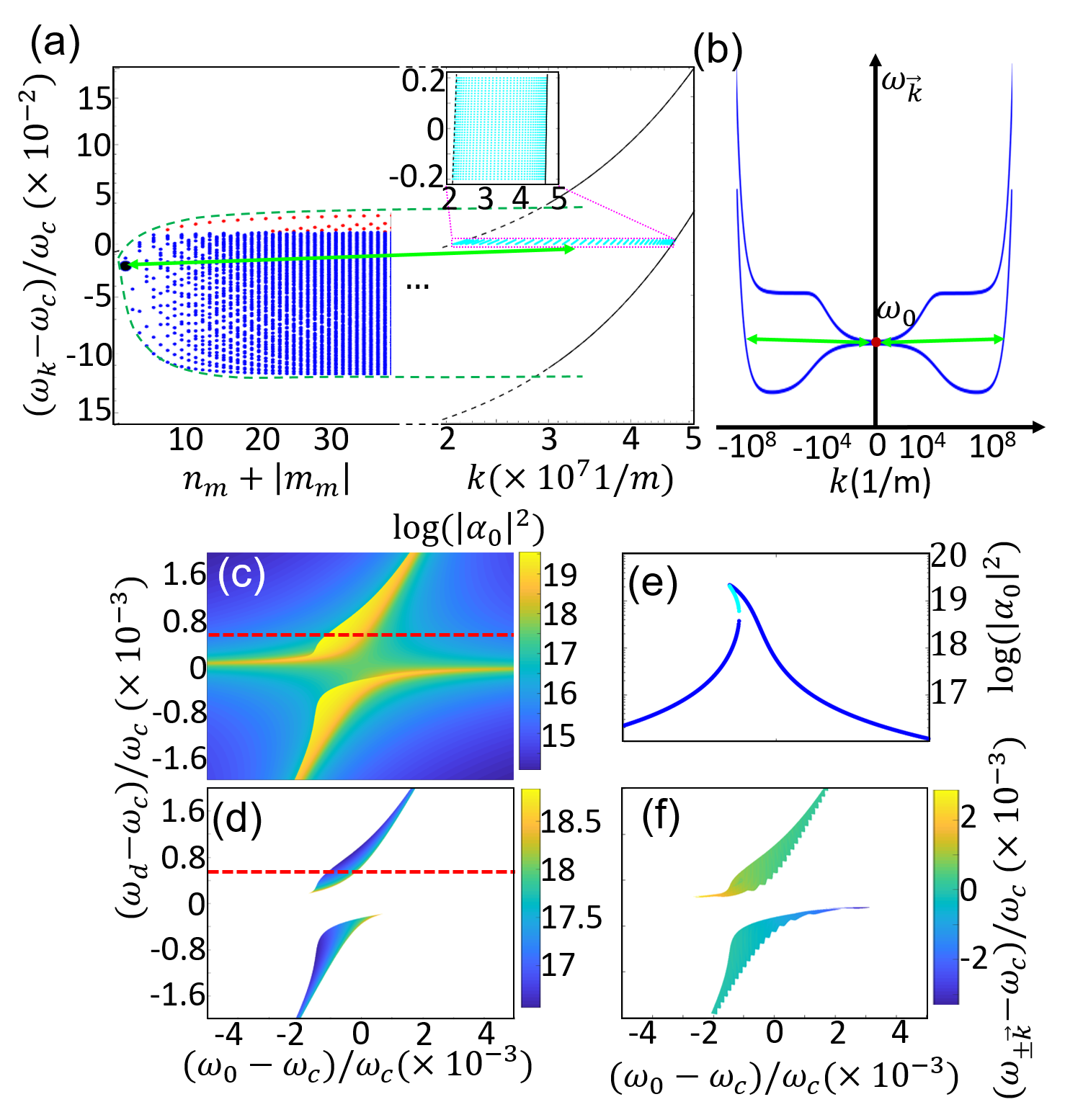}\caption{Kittel mode
instability in the Landau-Lifshitz-Gilbert equation. (a) The magnon dispersion
for a sphere. The magnetostatic (MS) modes are indicated by black (Kittel
mode), blue (bulk modes) and red (surface modes) dots. The MS band matches the
continuum band at certain $k$, at which the green dashed lines touch the black
full/dashed lines. The ellipsis $\cdots$ indicates the region that separates
the continuum from the MS manifold. The cyan-colored dots respresent the phase
space region tested for instabilities, of which the inset is a zoomed-in
version. (b) The envelopes of the magnon dispersion in a thin film on a
logarithmic wave number scale. The red dot indicates the Kittel mode. In (a) and (b) the top and bottom envelopes correspond to $\vec {k}\|\vec M_{0}\|\hat z$ and $\vec {k}\perp\vec M_{0}$, respectively. The green arrows in (a) and (b) schematically
depict the four magnon scattering processes (4MS) responsible for the instability of the Kittel mode. (c-e) Steady state solutions for the populated Kittel mode with
\textquotedblleft self-Kerr" nonlinearity but without inclusion of $\vec
{k}\neq0$ modes. (c) and (d): Magnon number $|\alpha_{0}|^{2}$ as a function
of Kittel mode and microwave drive frequency, where (c) depicts the solutions
when the system is in either stable state or the solution with larger magnon
population when the system is bistable, while (d) shows only the solution with
smaller magnon numbers in the bistable regime. (e) plots $|\alpha_{0}|^{2}$
along the red dashed lines of (c-d). Blue dots are the stable fixed points,
and the cyan dots are unstable saddle points. (f) Instability of the Kittel
mode solutions in (a) when $\vec{k}\neq0$ magnons are allowed to contribute.
Here the color codes the frequency of the $\pm\vec{k}$ pair that becomes
unstable first. }%
\label{fig1}%
\end{figure}

Strongly coupled MW photons can drive a weakly damped magnonic system easily
into the non-linear response regime. {Hysteresis
\cite{Bertotti2009,Stancil2009}, Bose-Einstein condensation
\cite{Demidov2007,Demidov2008,Bender2012,Serga2013,Bozhko2016},
auto-oscillation (and chaos)
\cite{Bryant1988,Rezende1990,Slavin2009,Demidov2012}, synchronization
\cite{Kaka2005,Elyasi2015}, soliton formation
\cite{Slavin1999,Buttner2000,Wu2006,Slavin2009,Demidov2012,Elyasi2019}, and
magnon transistors \cite{Chumak2014} are only few examples of non-linearities
in magnetism and magnonics.} Microwave cavities facilitate the study of
non-linear phenomena by focussing a large number of photons into narrow
frequency bands, for example leading to the observation of resonance frequency
shifts and bistability in an yttrium iron garnet (YIG) sphere as a function of
microwave intensity \cite{Wang2016_1,Wang2017}. These observations were
explained with the Duffing model --- the minimal model of a non-linear
oscillator, with an anharmonic term in the potential energy $\sim x^{4}$,
where $x$ is the canonical position. The Duffing model is the main means to describe non-linearities in the dynamics of nano- and
opto-mechanical systems
{\cite{Craighead2000,Aldridge2005,Shevchuk2015}}.

In the linear regime, the dynamics of the fundamental modes of optomechanics
and optomagnonics, such as the vibrations of a cantilever and the coherent
precession of the magnetic order (Kittel mode) obey basically the same
equations. However, while the Duffing model has been found to be quite appropriate for
most of non-linear mechanics, it is not obvious that it should work as well
for non-linear magnonics. For example, in contrast to the phonons in elastic
media, the magnetic dipolar interaction renders the magnon dispersion in thin
films strongly anisotropic and non-monotonic; the Kittel mode at the origin of
reciprocal space is not an energetic minimum. The three-field and four-field
magnon scattering processes caused by dipolar and exchange interactions, as
well as crystalline anisotropies, can lead to instabilities with finite wave
lengths that cannot be modelled by a single anharmonic oscillator. Indeed, the
unique spin wave dispersion is instrumental to some of the nonlinearity induced phenomena such as the generation and
observation of non-equilibrium Bose-Einstein (Rayleigh-Jeans) condensation of
magnons at nonzero wave vector
\cite{Demidov2007,Demidov2008,Serga2013,Ruckriegel2015,Bozhko2016}, magnonic
transistors \cite{Chumak2014}, and instabilities leading to classical
auto-oscillation and chaos \cite{Bryant1988,Rezende1990,Bertotti2009}.

On the other hand and in contrast to nanomechanics and Josephson devices,
magnonic quantum effects have been elusive with very few possible exceptions
\cite{Tabuchi2015,Lachance2017}. Observation of quantum non-linearities such
as squeezing, generation of non-classical states, photon blockade, and
entanglement
\cite{Aspelmeyer2014,Marquardt2006,Qian2012,Nunnenkamp2010,Rips2012,Wang2014}
have never been reported in magnonics.

Here we argue that transcending the Duffing paradigm is \textit{conditio sine
qua non} to explore a considerable potential of optomagnonics for quantum
applications. We start by discussing the magnon interactions of strongly
driven ferromagnets placed in microwave cavities. We discover not only interesting classical nonlinear dynamics of magnetization, but a variety of previously not investigated quantum effects and discuss how these effects can be used for the field of quantum information. By classical, semi-classical, and
quantum calculations we predict different interaction-induced classes of
steady states, including a genuinely quantum limit cycle of $\pm\vec{k}\neq0$
magnon modes. The fluctuation statistics of the steady states reveal
squeezing, which can serve as a quantum information resource, as well as
bounds for distillable entanglement and entanglement of formation. Finally, we
assess the effect of injection locking of the $\pm\vec{k}\neq0$ magnons limit
cycle on these entanglement measures. As few as four copies of steady states
can in principle be transformed into a completely entangled state equivalent
to a spin-singlet, which could be of immediate use in quantum teleportation,
simulation, and computation.

In Sec. \ref{sec1}, we introduce the details of the model of a magnet inside a
cavity and all the nonlinearities involved. Then we classify the outcomes of
the nonlinear terms in the anti-crossing region of the magnon Kittel mode and
cavity photon frequencies. The outcomes include bistability, Suhl instability,
fixed point, limit cycle, and chaotic dynamics of $\pm\vec{k}\neq0$ magnons. In
Sec. \ref{sec2}, we include quantum Langevin noise sources, and show that the
solutions of $\pm\vec{k}\neq0$ magnons excited by the Suhl instability of the
Kittel mode are always limit cycles. We also develop an equivalent quantum
master equation, and solve it in the number space of the corresponding
harmonic oscillators. The limit cycle of $\pm\vec{k}\neq0$ magnons is
reprodused in the Wigner function representation of the steady states. In Sec.
\ref{sec3}, we address the first quantum information resource, i.e.
fluctuation squeezing, which is observable by microwave scattering amplitudes.
In Sec. \ref{sec4}, we focus on entanglement as an important quantum
information resource. We find finite distillable entanglement shared between
the Kittel mode and $\pm\vec{k}\neq0$ modes. However, it is not simply
accessible via the covariance matrix of the (quantum) noise, and thereby less
interesting from an experimental point of view. In Sec. \ref{sec5_1}, we
introduce the mechanism of \textquotedblleft injection
locking\textquotedblright\ of the $\pm\vec{k}\neq0$ limit cycle solutions that
transforms an arbitrary phase excited state into a fixed point with Gaussian
statistics. In Sec. \ref{sec5_2}, we show that the distillable entanglement
then becomes accessible in the covariance matrix, allowing for a
straightforward experimental analysis and utilization of the entanglement. In
Sec. \ref{sec5_3}, we assess the effect of injection locking on entanglement
calculated from the quantum master equation solution, and analyze the
consistency with the semclassical approaches of Sections \ref{sec5_1} and
\ref{sec5_2}. Finally, in Sec. \ref{sec6}, we propose concrete set-ups for
experimental realization of the quantum information resources assessed in this
work, addressing the key parameters, feasibility, challenges, and constraints
on e.g. magnet dimension and environment temperature.

\section{Model and classical nonlinear analysis}

\label{sec1} We focus (but do not limit) attention on a high-quality magnetic
element such as a sphere (or cube) of yttrium iron garnet in a microwave
cavity. The static magnetization $\vec{M}_{0}$ is saturated and aligned by an
applied static magnetic field $\vec{H}_{ext}\Vert\hat{z}.$ The magnet is
placed into the antinode of a transverse AC magnetic field of a selected
cavity mode with angular frequency $\omega_{c}$. In the total Hamiltonian
\begin{equation}
H^{(T)}=H^{(c)}+H^{(mc)}+H^{(d)}+H^{(T,m)}, \label{eq1}%
\end{equation}
$H^{(c)}=\hbar\omega_{c}b^{\dag}b$, where $b$ ($b^{\dag}$) is the annihilation
(creation) operator of a bare photon cavity mode, respectively. $H^{(mc)}$ is
the magnon-photon interaction, $H^{(d)}=i\bar{B}\left(  e^{-i\omega_{d}%
t}b^{\dag}-e^{i\omega_{d}t}b\right)  $ is the microwave input drive with
frequency $\omega_{d}$ and amplitude $\bar{B}$. $H^{(T,m)}=\sum_{\vec{k}}%
\hbar\omega_{\vec{k}}c_{\vec{k}}^{\dag}c_{\vec{k}}+H_{int}^{(T,m)}$ governs
the magnons with annihilation/creation field operators $c_{\vec{k}}/c_{\vec
{k}}^{\dag}$. The dispersion relation $\omega_{\vec{k}}$ and their
interactions $H_{int}^{(T,m)}$ are affected by dipolar field, exchange
interaction, and crystalline anisotropy, as summarized in Appendix \ref{app1}
\cite{Bryant1988,White2006,Rezende2009}. Magnons in the bulk of a magnet are
plane waves with wave vector $\vec{k}$ and frequency $\omega_{\vec{k}}$ that
start from the Kittel mode at $\vec{k}=0$ and can be very anisotropic [see
Figs. \ref{fig1}(a) and (b)] \cite{Walker1957,Walker1958,Damon1961,Kalinikos1986,Hurben1995}. We treat the finite size effects at wavelengths
comparable to the sample size in the Suhl approximation \cite{Suhl1957},
separating the uniform mode from those with finite wavelength that we treat as
plane waves. At long wave lengths we may invoke the magnetostatic (MS)
approximation \cite{Walker1957,Walker1958} to treat the effects of the dipolar
interaction. We focus on YIG spheres that are often used in experiments
\cite{Zhang2014,Tabuchi2015}, but we can handle magnetic films with minor adjustments.

\begin{figure}[!]
\includegraphics[width=0.5\textwidth]{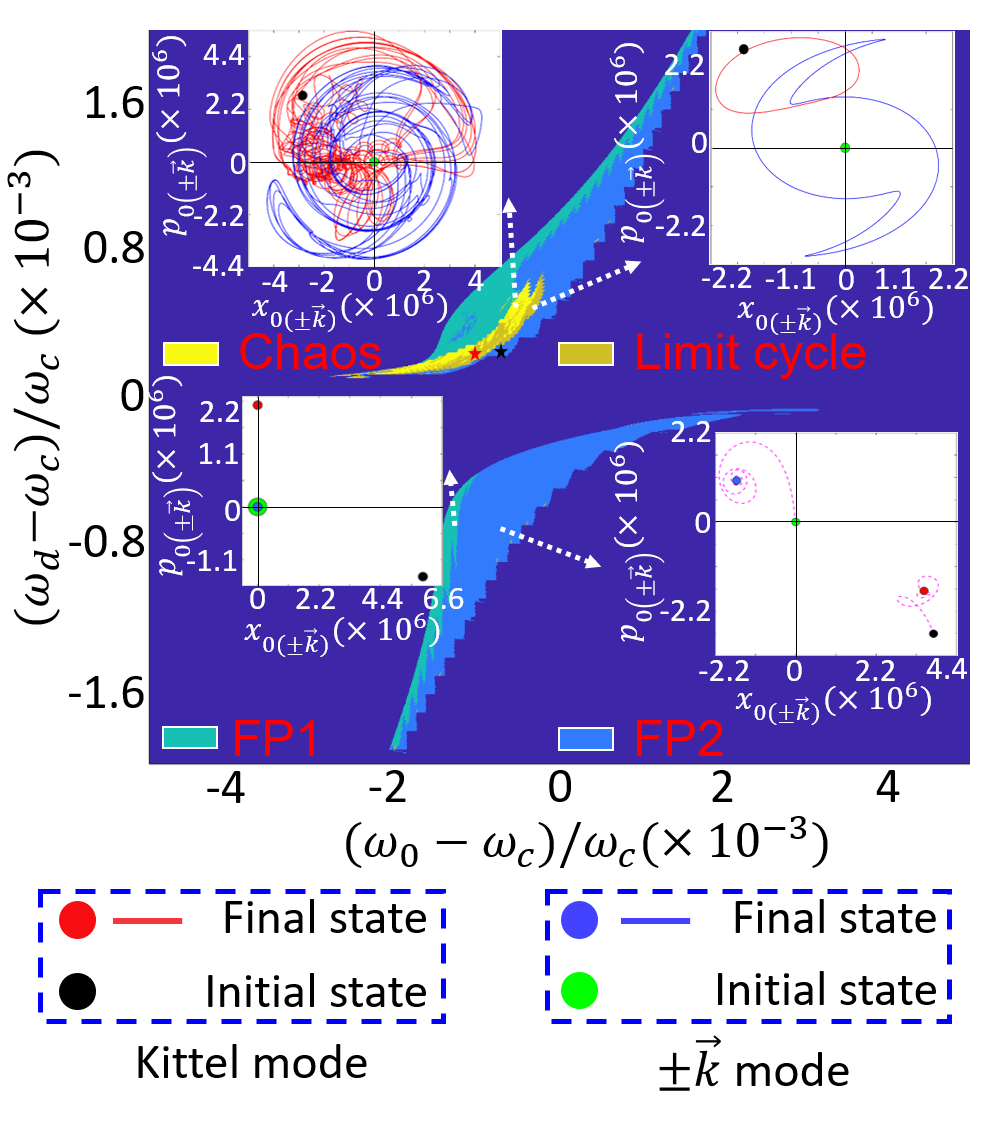}\caption{Steady state
classification of the solutions of the Landau-Lifshitz equation without
thermal noise. The darkest blue indicates the region where there is no
instability of the Kittel mode to $\pm\vec{k}\neq0$. Each type of steady state
is indicated by a color labelled inside the main panel, i.e. chaos, limit
cycle, FP1, and FP2. For each of the four types, an example is also shown, and
the corresponding point in the main panel map is indicated by white dashed
arrows. Point and line colors in inside panels corresponding to initial states
and final states are shown in the bottom panel. In FP1 inset, the green point
size is adjusted for clarity, and the blue dot coincides with the green dot. The
trajectory from initial to final state of FP2 is also shown. The values
corresponding to the photonic mode are not shown. The black and red stars in
the figure are the $(\omega_{0},\omega_{d})$ values used in Fig. \ref{fig3}.}%
\label{fig2}%
\end{figure}

YIG spheres can be fabricated with diameters down to 250 $\mathrm{\mu}$m with
traditional technology \cite{Zhang2015,Morris2017}. By integrating YIG into
nanoscale photonic chips, further downscaling appears possible
\cite{Morris2017,Zhu2017,*Heyroth2019}. Here, for calculations, we chose a diameter of 0.1$\,$mm,
keeping in mind that for larger (smaller) spheres, larger (smaller) drive
powers are required to achieve the same results. We adopt a saturation
magnetization $M_{s}=1.46\times10^{5}\,\mathrm{A/m}$ along the $(111)$
crystalline axis with uniaxial magnetic anisotropy $K_{c}=-2480\,\text{J/m}%
^{3}$\cite{Hansen1974}, and gyromagnetic ratio $\gamma=2.11\times
10^{5}\,\mathrm{m/(As)}$ \cite{Stancil2009}. We adopt a cavity mode
$\omega_{c}=10^{11}/(2\pi)\,\text{1/s}$ and Kittel-cavity mode
coupling constant $D_{0}=10\,\text{MHz}$. The magnetization dynamics is damped
by a Gilbert constant $\alpha_{\text{G}}=10^{-4}$ \cite{Chang2014} and the
dissipation rate of the cavity mode $\zeta_{c}=1\,\text{MHz}$
\cite{Tabuchi2015}. The magnon interaction in a gas with finite density is
described by the Holstein-Primakoff expansion (see Appendix \ref{app1}) in
terms of crystalline anisotropies, dipolar and exchange interactions. The
leading 3 and 4 particle magnon-magnon scattering processes are conveniently treated in
a (Suhl) compartmentalized reciprocal space. The magnon-magnon scattering terms for the
Kittel mode are e.g. $c_{0}^{\dag}c_{-\vec{k}}c_{\vec{k}}$ and $c_{0}^{\dag
}c_{0}^{\dag}c_{-\vec{k}}c_{\vec{k}}$, where $c_{0(\vec{k})}^{\dag}%
$/$c_{0(\vec{k})}$ are the creation/annihilation operators of the Kittel and
plane-wave $(\vec{k}\neq0)$ magnon modes, respectively. Energy and momentum
conservation impose constraints on the ($\vec{k}\neq0,\omega_{\vec{k}}$)
states into which a Kittel mode magnon can be scattered. We assume an external
magnetic field $\vec{H}_{ext}$ large enough (much greater than $M_{s}/3$ for a sphere) such
that three magnon, as well as two magnon-one photon scattering processes are
non-resonant, which allow us to focus on the effects of four-magnon scatterings (4MS). The degeneracy we exploit here vanishes for
very small magnets, so the particle diameter $d\gtrsim
1\,\mathrm{\mu}$m\textit{.}

As discussed in Appendix \ref{app1}, essentially all the dipolar
interaction, exchange, and anisotropy contribute to the 4MS terms, see e.g.
Eqs. (\ref{eq_a10}-\ref{eq_a14}). Here, the exchange interaction maximizes the coefficient of the 4MS
term responsible for Suhl instability at large $|\vec{k}|$. The shape
anisotropy and crystalline anisotropy contribute a repulsive 4MS interaction.
Dipolar interaction and crystalline anisotropy mix the modes, leading to
complex 4MS terms [see Eqs. (\ref{eq_a18}-\ref{eq_a20})] such that the label attractive or repulsive cannot be simply
made. While the formalism is material independent, we focus here on a
parameter set for undoped YIG materials.

The field operators $c_{0}=\alpha_{0}+\delta c_{0}$, $c_{\vec{k}}=\alpha
_{\vec{k}}+\delta c_{\vec{k}}$ and $b=\beta+\delta b$ fluctuate by $\{\delta
c_{0},\delta c_{\vec{k}},\delta c_{0}\}$ around the steady state mean field
values $\{\alpha_{0},\alpha_{\vec{k}},\beta\}$. The canonical position and
momentum (quadrature operators) are $x_{0(\vec{k})}=(c_{0(\vec{k})}^{\dag
}+c_{0(\vec{k})})/2$ and $p_{0(\vec{k})}=i(c_{0(\vec{k})}^{\dag}-c_{0(\vec
{k})})/2$ for the magnon modes and $X=(b^{\dag}+b)/2$ and $Y=i(b^{\dag}-b)/2$
for the photon mode, respectively, with fluctuations $\delta x_{0(\vec{k})}$,
$\delta p_{0(\vec{k})}$, $\delta X$, and $\delta Y$, respectively.

We present results in the $(\omega_{0},\omega_{d})$ parameter space for
$\bar{B}=3.3\times10^{13}\,\mathrm{s}^{-1}$ corresponding to $\sim1\,\mathrm{mT}$ order of magnitude for dynamic magnetic field exerted on the magnet when $\omega_{d}\sim\omega
_{c}$. $\bar{B}=\sqrt{\zeta_{c,{ex}}P_{in}/(\hbar\omega_d)}$, where $\zeta_{c,{ex}}$ is the photon dissipation by leakage, and the cavity input power
$P_{in}\approx13\,$m$\text{W}$ ($11.1\,\text{dBm}$) for the range of $\omega_d$ we consider. This power should be large enough to
access all phenoma offered in nonlinear phase space. For smaller powers down
to $P_{in}\sim1\,$mW, we observe bistability of the Kittel mode and the Suhl
instability close to the origing of $(\omega_{0},\omega_{d})$ parameter space,
but no limit cycles and/or chaotic motion. The power
demands scale with the volume of magnet: for a sphere of $1\,$mm radius, a
$P_{in}=13\,$W is necessary to achieve the same results as shown here. Bryant et al. {\cite{Bryant1988}}
carried out a classical analysis of the full nonlinear dynamics of the Suhl
instability of the first kind as a function of input power and dc external
magnetic field, at twice the microwave frequency of the Kittel mode. T{he
nonlinear magnetization dynamics of YIG spheres with a typical diameter of
$1\,$mm required powers in the range of $1-20\,$dB ($1-100\,$W)
\cite{Bryant1988,Rezende1990}. }Here we address the Suhl instability of the
second kind at excitation frequencies close to the Kittel mode. Experiments
that classified the corresponding nonlinear phase space observed limit cycles,
and their doublings, eventually leading to chaos \cite{Gibson1984}. The
microwave magnetic fields that cause classical chaos in our analysis is
$\sim2\,$mT when $\omega_{0}\sim\omega_{d}$, which agrees with the
$\sim1-10\,$mT [$\sim10$ times critical field of Suhl instability
\cite{Suhl1957,Naletov2007} of the second kind] in these experiments and theories
\cite{Gibson1984,Rezende1990}.

The equations of motion (EOM) for the three distinct fields (and similarly for the hermitian conjugates), viz. the cavity
mode field, Kittel mode field, and selected $\vec{k}\neq0$ modes fields, is
obtained from $H^{(T)}$ [see Eq. (\ref{eq1}) and Appendix \ref{app1}], with
dissipation and noise added:
\begin{align}
\dot{c}_{0}  &  =-i\left(  \Delta_{0}+2\sum_{\vec{k}\neq0}\operatorname{Re}%
\left[  \mathcal{D}_{0,\vec{k}}^{4MS,1}\right]  n_{\vec{k}}\right)
c_{0}-\frac{\zeta_{m,0}}{2}c_{0}\nonumber\\
&  -2ic_{0}^{\dag}\sum_{\vec{k}\neq0}\mathcal{D}_{0,\vec{k}}^{4MS,2}%
c_{-\vec{k}}c_{\vec{k}}-\nonumber\\
&  2i\operatorname{Re}[\mathcal{D}_{0,0}^{4MS,1}+\mathcal{D}_{0,0}%
^{4MS,2}][c_{0}+2c_{0}^{\dag}c_{0}c_{0}]+iD_{0}b+\nonumber\\
&  \sqrt{\zeta_{mm,0}}F_{mm,0}\left(  t\right)  +\sqrt{\zeta_{mp,0}}%
F_{mp,0}\left(  t\right)  , \label{eq34_new}%
\end{align}%
\begin{align}
\dot{c}_{\vec{k}\neq0}  &  =-i\left(  \Delta_{k}+2\sum_{\vec{k}^{\prime}%
\neq\vec{k}}\operatorname{Re}\left[  \mathcal{D}_{\vec{k},\vec{k}^{\prime}%
}^{4MS,1}\right]  n_{\vec{k}^{\prime}}\right)  c_{\vec{k}}-\frac{\zeta
_{m,\vec{k}}}{2}c_{\vec{k}}\nonumber\\
&  -i(\mathcal{D}_{0,\vec{k}}^{4MS,2})^{\ast}c_{0}c_{0}c_{-\vec{k}}^{\dag
}-2i\operatorname{Re}\left[  \mathcal{D}_{\vec{k},\vec{k}}^{4MS,1}\right]
\left[  c_{\vec{k}}+2c_{\vec{k}}^{\dag}c_{\vec{k}}c_{\vec{k}}\right]
\nonumber\\
&  -ic_{-\vec{k}}^{\dag}\sum_{\vec{k}^{\prime}\neq0}\mathcal{D}_{\vec{k}%
,\vec{k}^{\prime}}^{4MS,2}c_{-\vec{k}^{\prime}}c_{\vec{k}^{\prime}}\nonumber\\
&  +\sqrt{\zeta_{mm,\vec{k}}}F_{mm,\vec{k}}\left(  t\right)  +\sqrt
{\zeta_{mp,\vec{k}}}F_{mp,\vec{k}}\left(  t\right)  , \label{eq35_new}%
\end{align}%
\begin{align}
\dot{b}  &  =-i\Delta b-\frac{\zeta_{c}}{2}b+B+iD_{0}c_{0}+\nonumber\\
&  \sqrt{\zeta_{c,0}}F_{c,0}(t)+\sqrt{\zeta_{c,ex}}F_{c,ex}(t),
\label{eq36_new}%
\end{align}
where $\mathcal{D}_{\vec{k}^{\prime},\vec{k}^{\prime\prime}}^{4MS,1}$ and
$\mathcal{D}_{\vec{k}^{\prime},\vec{k}^{\prime\prime}}^{4MS,2}$ are the
strengths of the 4MS scatterings of the form $c_{\vec{k}^{\prime}}^{\dag
}c_{\vec{k}^{\prime}}c_{\vec{k}^{\prime\prime}}^{\dag}c_{\vec{k}^{\prime
\prime}}$ and $c_{\vec{k}^{\prime}}^{\dag}c_{-\vec{k}^{\prime}}^{\dag}%
c_{\vec{k}^{\prime\prime}}c_{-\vec{k}^{\prime\prime}}$, respectively (see
Appendix \ref{app1}). We defined detunings $\Delta_{0}=\omega_{0}-\omega_{d}$,
$\Delta_{\vec{k}}=\omega_{\vec{k}}-\omega_{d}$, and $\Delta=\omega_{c}%
-\omega_{d}$. The damping parameters are $\zeta_{m,0}=\zeta_{mm,0}%
+\zeta_{mp,0}$, $\zeta_{m,\vec{k}}=\zeta_{mm,\vec{k}}+\zeta_{mp,\vec{k}}$, and
$\zeta_{c}=\zeta_{c,ex}+\zeta_{c,0}$. $\zeta_{mm,0}$ ($\zeta_{mp,0}$) is the
dissipation rate of the Kittel mode field by interaction with the magnon
(phonon) bath, and $\zeta_{mm,\vec{k}}$ ($\zeta_{mm,\vec{k}}$) the same for
the $\vec{k}$ mode. $\zeta_{c,ex}$ ($\zeta_{c,0})$ is the photon dissipation
by leakage (interaction). The damping parameters $\zeta_{X}$ are connected to
the stochastic (Markovian) Langevin fields $F_{X}$ by the fluctuation
dissipation theorem. Details are given in Appendix \ref{app2}.

The nonlinearity of the Kittel mode alone is the so-called self-Kerr term
$(c_{0}^{\dag}c_{0})^{2},$ which corresponds to the non-parabolicity in the
Duffing model. It leads to a bistability in the solutions of the classical
mean fields when $\bar{B}$ exceeds a certain threshold, which happens here in
$(\omega_{0},\omega_{d})$ parameter space close to the magnon-polariton
$(\omega_{0}=\omega_{d})$ \cite{Drummond1980,Wang2016_1,Wang2017} We address
this reduced problem by the EOM of Eqs. (\ref{eq34_new}-\ref{eq36_new}), by
dropping all terms involving the $\vec{k}\neq0$ magnons and replace the
stochastic fields by their mean values, which leads to a sixth order equation
in $|\alpha_{0}|^{2}$ ($\alpha_0$ is the Kittel mode mean field).
Our choice for\textit{ }$\bar{B}$ is above the threshold, and two stable and
one unstable (saddle point) solutions of $|\alpha_{0}|^{2}$ manifest the
classical bistability. Figures \ref{fig1}(c)-(e) show a typical map
of the computed \textquotedblleft self-Kerr\textquotedblright\ solution, i.e.
the Kittel magnon number $|\alpha_{0}|^{2}$ without mixing with other modes.
Figure \ref{fig1}(c) summarizes the stable solutions and the large amplitude
or number state in the parameter regime in which the system is bistable, i.e.
close to the magnon-polariton (anti)crossing, while Figure \ref{fig1}(d) is
the other stable solution with smaller magnon numbers. Note that
we evaluate the complete nonlinear phase space at each $(\omega_0,\omega_d)$ independently, thereby
disregarding (classical) hysteretic effects that arise when cycling e.g. the applied magnetic
field (i.e. $\omega_0$). Figure \ref{fig1}(e) is a plot of the frequencies of the stable and
unstable (saddle points) solutions for the $\omega_{d}$ indicated by
red-dashed lines in Figs. \ref{fig1}(c) and \ref{fig1}(d).

Subsequently, we assess the Suhl instability of the solutions in Fig.
\ref{fig1}(c) caused by 4MS with $\vec{k}\neq0$ magnons (for the parameters
used here, the lower magnon number solutions in Fig. \ref{fig1}(d) remain
stable). We scan the $\vec{k}$ values for which the 4MS is expected to be
largest (see Fig. \ref{fig1}(a)) and search for the $\pm\vec{k}\neq0$ pair of
modes with largest positive real eigenvalue of the linearized matrix
$\mathcal{O}$ (defined in Appendix \ref{app2}) that here corresponds to the EOM linearized around the Kittel mode mean field. Results are summarized in Fig. \ref{fig1}(f)
in the form of the frequencies of the most unstable magnon pairs $\omega
_{\pm\vec{k}}$.

In order to classify the steady states, we first solve the EOM, i.e. Eqs.
(\ref{eq34_new}-\ref{eq36_new}) without Langevin stochastic fields, in which
the field operators become classical amplitudes, similar to conventional
micromagnetics. The solutions of the EOM with initial condition chosen to be
an excited pure Kittel mode from Fig.\textit{ }\ref{fig1}(c) are shown in
Figure \ref{fig2}. We observe (i) chaotic behavior with finite positive
Lyapunov exponents \cite{Eckmann1985,Bryant1990}, (ii) limit cycle (LC), and
(iii) fixed point (FP1 and FP2) solutions. The final state in region FP1 is a
pure Kittel mode, which implies that the self-Kerr solution from the higher
magnon number branch relaxes back to the stable lower magnon number one in
Fig. \ref{fig1}(d): The $\vec{k}\neq0$ modes help the Kittel mode to explore
a larger phase space, thereby escaping a fixed point with a shallow energy
well. In region FP2 the system settles into a hybrid state with significant
contributions from magnons with finite momentum.

\begin{figure}[!]
\includegraphics[width=0.5\textwidth]{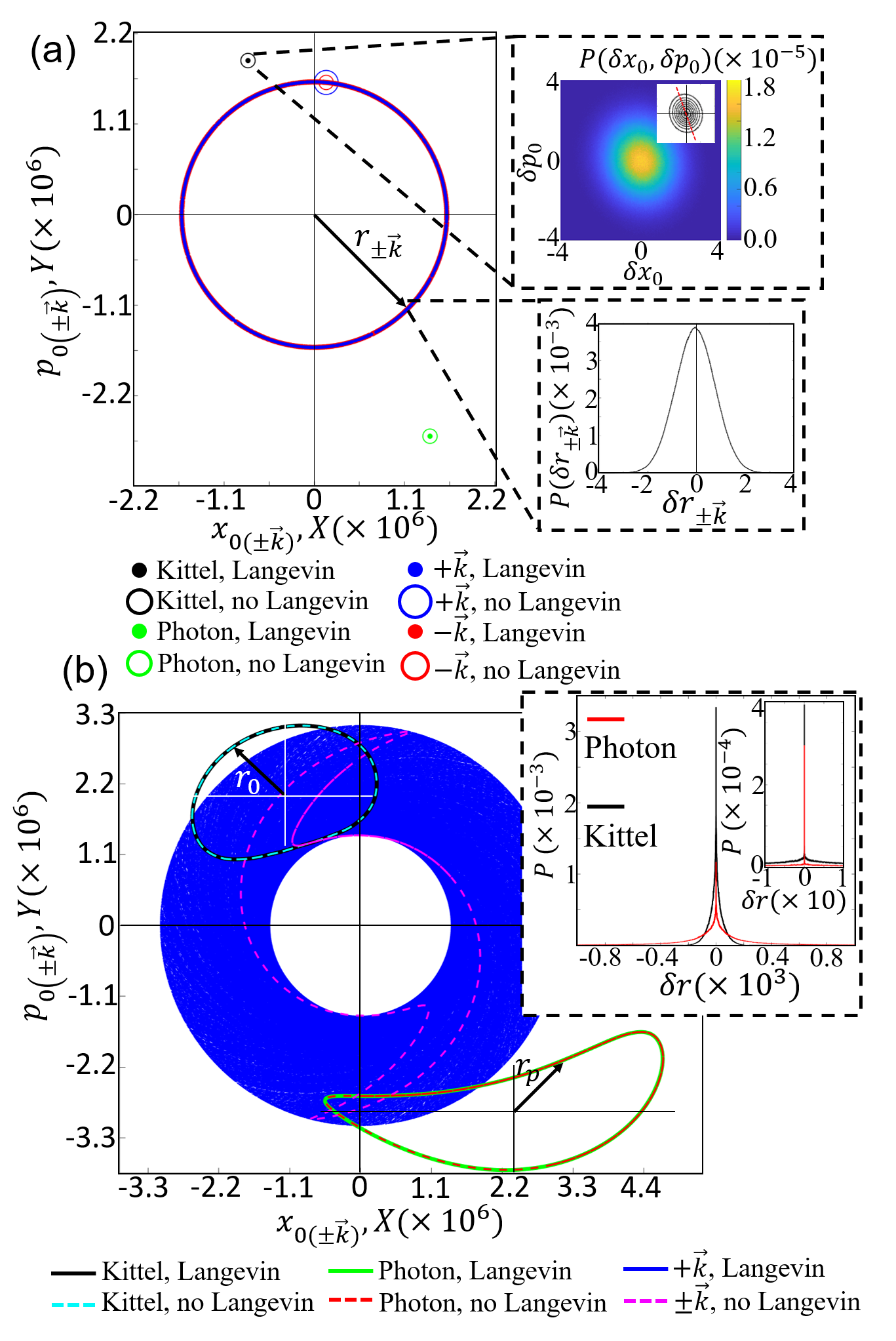}\caption{Steady states of the system with
inclusion of Langevin fields. (a) and (b) correspond to FP2 and LC, at
$(\omega_{0},\omega_{d})$ indicated by black and red stars in Fig. \ref{fig2},
respectively. (a) Main panel is the final fixed points with and without
inclusion of Langevin fields for the Kittel, $\pm\vec{k}$, and photon modes.
The final states for all the 1600 runs with Langevin fields, plotted. Insets
show probability distributions of the fluctuations. In the panel for $P(\delta
x_{0},\delta p_{0})$, the inset is contour plot, and the red dashed line
indicates the long axis of the fluctuation ellipse. (b) Main panel shows the
trajectory of all the modes in the last $2\,\mu\text{s}$ of all the 1600 runs
with Langevin fields. The trajectories of steady states without Langevin
fields also shown. The inset shows the probability distribution of photon and
Kittel modes. In (a) and (b), the probability distribution is evaluated over
the fluctuations in the width of the limit cycles, i.e. $\delta r_{\pm\vec{k}%
}$, $\delta r_{0}$, and $\delta r_{p}$, averaged over the cycle loop. In each
panel, $r_{\pm\vec{k}}$, $r_{0}$, and $r_{p}$ indicate $\pm\vec{k}$, Kittel,
and photon modes limit cycles with respect to their corresponding centers.
Centers of each limit cycle is crossing of global/local axes shown.}%
\label{fig3}%
\end{figure}
\begin{figure*}[!]
\includegraphics[width=0.9\textwidth]{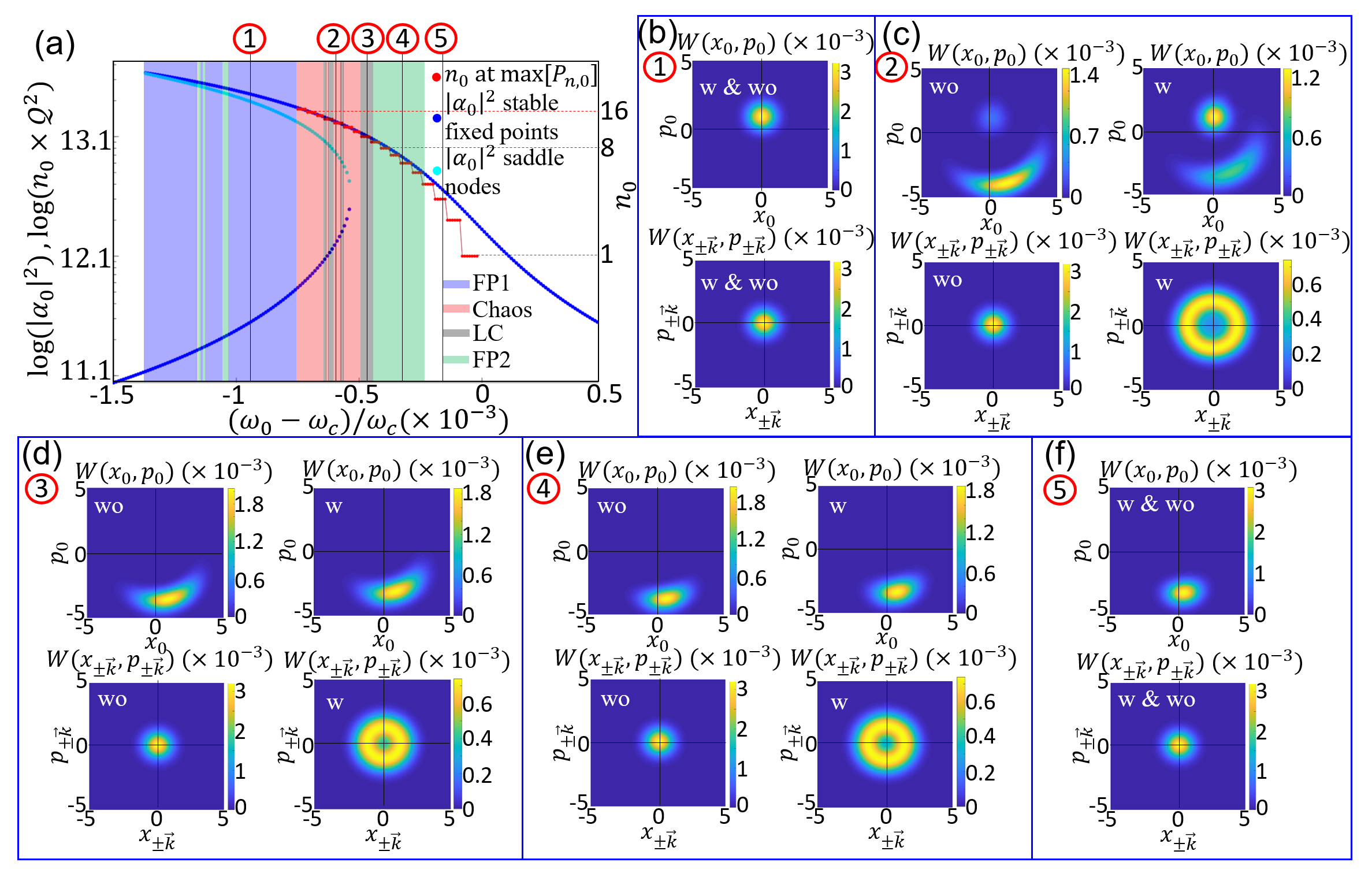}\caption{Steady states
derived from the quantum master equation. (a) Duffing model: $n_{0}$, the
maximum of the number distribution $P_{n,0}$ of the Kittel mode compared with
the classical magnon number ($|\alpha_{0}|^{2}$) from Fig. \ref{fig1}(e) for
the photon frequency $\omega_{d},$ indicated by the red-dashed line in Figs.
\ref{fig1}(c) and (d). The right vertical ordinate indicates $n_{0}$ as
solved in the scaled system, whereas the left axis shows the corresponding
values in the physical basis. The different classes of steady states when
$\vec{k}\neq0$ modes are included are coded by the background shading colors
(see Fig. \ref{fig2}). Five particular magnetic fields ($\omega_{0})$ are
singled out by the numbered vertical thin black lines. (b)-(f) Wigner function
[Eq. (\ref{eq2})] for the Kittel and $\pm\vec{k}$ modes [`w'(`wo') means
with(out) inclusion of $\pm{\vec{k}}$ modes].}%
\label{fig4}%
\end{figure*}

\section{Quantum Langevin and master equation}

\label{sec2} Next, we solve the EOMs with random initial conditions close to
those of Fig. (\ref{fig2}) and add Langevin quantum stochastic fields at an
ambient temperature $T_{\mathrm{env}}=1\,\text{K}$ (see Eqs. (\ref{eq34_new}%
-\ref{eq36_new}) and Appendix \ref{app2_2} for description of baths and their
correlation functions, moderate temperature variations cause expected and mild
changes). Since the (Markovian) bath approximation breaks down as
$T_{\mathrm{env}}\rightarrow0,$ we cover the ultra-low temperature regime by
solving the quantum master equations for $T=0\,$K (see also Appendix
\ref{app3}). We conclude below that the features such as entanglement measures
(see Sec. \ref{sec5}) are mathematically (and physically) tolerant with
respect to moderate changes in temperature. We repeat the computations 1600
times for the $(\omega_{0},\omega_{d})$ points of Figure \ref{fig2} indicated by black and red stars, i.e. a
fixed point of type 2 (FP2), with results in Figure \ref{fig3}(a), and of a
limit cycle (LC) with results in Figure \ref{fig3}(b), respectively. We plot the probability
distributions of $\{\langle x_{0}(p_{0})\rangle,\langle X(Y)\rangle,\langle
x_{\pm\vec{k}}^{2}+p_{\pm\vec{k}}^{2}\rangle\}$, where $x_{0}%
=\mathit{\operatorname{Re}}\left[  \alpha_{0}\right]  $, $p_{0}%
=\operatorname{Im}\left[  \alpha_{0}\right]  $ for the Kittel mode,\textit{
}$X,Y$ are the analogues for the photon (for completeness),\textit{ }and
$\left\vert \alpha_{\pm\vec{k}}\right\vert ^{2}=x_{\pm\vec{k}}^{2}+p_{\pm
\vec{k}}^{2}$ is the number of $\pm\vec{k}$ magnons\textit{.} The phase of
$\alpha_{\pm\vec{k}}$ for FP2 in Figure \ref{fig3}(a) becomes undetermined.
Even though the steady state solution of $\vec{k}\neq0$ magnons is a fixed
point, the noise transforms it to a limit cycle. The Kittel and photon modes,
on the other hand, undergo only a coherent precession with small and
elliptical fluctuations while their\textit{ }phases remain
deterministic.\textit{ }Figure \ref{fig3}(b) shows the fate of a LC after the
noise is switched on. The averages of the $\vec{k}\neq0$ magnons are
distributed over a doughnut in phase space [a typical trajectory is plotted as
pink dashed line in Figure \ref{fig3}(b)]. The dynamics of the Kittel and
photon mode are still the same deterministic LC closed loops of the noiseless
solutions. The inset of Fig. \ref{fig3}(a) shows the probability distribution
$P$ of the fluctuations around the mean field of the Kittel and $\pm\vec{k}$
modes. Since the mean-field Kittel (or photon) mode is a fixed point, the
fluctuation probability distribution in $(\delta x_{0},\delta p_{0})$ space is
expected to be Gaussian. The radial fluctuations $P(\delta r_{\pm\vec{k}}%
)\ $of the $\pm\vec{k}$ limit cycle solutions are also Gaussian distributed.
On the other hand, the fluctuations around the Kittel (photon) mode limit
cycles in the inset of Fig. \ref{fig3}(b) are clearly not Gaussian. The width
of the distribution is larger by a factor of $\sim10^{3}$ than the Gaussians
of Fig. \ref{fig3}(a) that correspond to the environment temperature
$T_{env}=1\,\text{K.}$ The probability distribution is the deviation from the
trajectory, and in principle independent of its form. The relatively large
width of the distribution appears nevertheless to be correlated with the
complexity of the deterministic LC, and the increased noise smear out the
structure in phase space. We see below that these LC's do not exist in the
quantum master equation calculations, indicating that the quantum fluctuations
have more serious effects than the thermal ones at 1 K.

Finally, we turn to quantum effects by solving the master equation
\begin{equation}
\dot{\hat{\rho}}=-i[H^{\prime(T)},\hat{\rho}]+\sum_{\vec{k}^{\prime}%
\in\{0,\vec{k},-\vec{k}\}}L_{\vec{k}^{\prime}}^{(T)}(\hat{\rho}%
,T_{\mathrm{env}}), \label{eq2_new}%
\end{equation}
where in $H^{\prime(T)}$ the photon mode has been adiabatically removed from
$H^{(T)}$, with renormalized Kittel mode detuning $\Delta_{0}%
^{\prime}=[\Delta_{0}-(D_{0}^{2}\Delta)/(\Delta^{2}+\zeta_{c}^{2}/4)]$ and
Kittel mode drive $i(\bar{B}^{\prime}c_{0}^{\dag}-h.c.)$ with effective field
$\bar{B}^{\prime}=(-i\Delta\bar{B}D_{0})/(\Delta^{2}+\zeta_{c}^{2}/4)$, where
$\Delta$ is cavity detuning with respect to drive and $\zeta_{c}$ the cavity
damping. $L_{\vec{k}^{\prime}}^{(T)}(\hat{\rho},T_{\mathrm{env}})$ is the
total Lindblad operator for each of the magnon modes. This master equation is
based on the assumption of Markovian baths. Equation (\ref{eq2_new}) can be
rewritten in terms of a super-operator $\mathcal{L}$ as $\dot{\hat{\rho}%
}=\mathcal{L}\rho$ and the steady state density matrix $\rho_{ss}$ is the
solution of $\mathcal{L}\rho_{ss}=0$. We can compute the eigenvector of the
sparse matrix $\mathcal{L}$ corresponding to the lowest eigenvalue (see
Appendix \ref{app3}) for a matrix dimension of up to $10^{6}\times10^{6}$.
Even at $T_{\mathrm{env}}=0$ this forces us to scale the system down to a
numerically tractable Hilbert space, dividing $\bar{B}$ by a factor
$\mathcal{Q}$ while multiplying the fourth order interactions by
$\mathcal{Q}^{2}$. This scaling preserves the bistability map as well as
instability with respect to $\pm\vec{k}$ magnon generation in the $(\omega
_{0},\omega_{d})$ parameter space. The cost-benefit ratio of the scaling is
optimized by $\mathcal{Q}=1.1\times10^{6}$. Calculations for finite
temperatures are possible but expensive, and for this scaling amplitude, we
expect only weak effects for $T_{\mathrm{env}}\lesssim0.1\,\text{K.}$

We first focus on the quantum mechanical Duffing oscillator, without mixing in
$\pm\vec{k}$ magnons. Figure \ref{fig4}(a) summarizes the calculated
$|\alpha_{0}|^{2}$, as well as the $n_{0}=c_{0}^{\dag}c_{0}$ which maximizes
the number distribution for a fixed photon frequency, and $\omega_{d}$
corresponding to Fig. \ref{fig1}(e). The left axis shows $|\alpha_{0}|^{2}$
and the rescaled $n_{0}\mathcal{Q}^{2}$ to facilitate comparison with the
classical results, while the right axis is $n_{0}$ in the downscaled system.
The discrete steps in $n_{0}$ are an artifact introduced by the small size of
the rescaled system. The colored background encodes the type of the
corresponding classical steady state. The numbered vertical lines indicate
the selected drive frequencies for which we compute the steady state density
matrix $\rho_{ss}$ including the $\vec{k}\neq0$ magnon modes. The reduced
density matrix for each mode $\rho_{\vec{q}}$ is obtained by tracing out all
other modes, i.e. $\rho_{\vec{q}}=\mathrm{Tr}_{\vec{q}^{\prime}\neq\vec{q}%
}\left[  \rho_{ss}\right]  ,$ where $\vec{q}^{\prime},\vec{q}\in\left\{
0\text{ (Kittel magnon),}\pm\vec{k}\right\}  $. We calculate the Wigner
function, a (quasi-)probability distribution in position-momentum phase space
\cite{Carmichael1999,Walls2008},
\begin{equation}
W(x_{\vec{q}},p_{\vec{q}})=\int\left\langle x_{\vec{q}}-\frac{y}{2}\right\vert
\hat{\rho}_{\vec{q}}\left\vert x_{\vec{q}}+\frac{y}{2}\right\rangle
e^{ip_{\vec{q}}y}dy, \label{eq2}%
\end{equation}
where $\left\vert x_{\vec{q}}\pm\frac{y}{2}\right\rangle $ are position
eigenstates. Results are summarized in Figures \ref{fig4}(b)-(f), representing
the distinct classes (FP1, chaos, LC, FP2, and stable to 4MS, respectively)
found in the (semi-)classical calculations. While chaos and limit cycles of
the Kittel and photon modes do not survive in the quantum regime, the limit
cycle in the $\pm\vec{k}\neq0$ modes become conspicuous as rings in Figs.
\ref{fig4}(c)-(e).The maxima of the Wigner functions should be interpreted as
attractors (fixed point or limit cycle) that are broadened by zero-point (and
in case of $T_{env}\neq0$ thermal) fluctuations. A fixed point with
(squeezed) thermal fluctuations such as the Kittel mode solution in Fig.
\ref{fig3}(a) becomes a Gaussian in the Wigner function centered on the same
point of phase space, e.g.
\begin{equation}
W(x_{0},p_{0})=\frac{1}{\pi(n_{th}+\frac{1}{2})}\exp\left(  {-\frac{\left\vert
x_{0}+ip_{0}-\alpha_{0}\right\vert ^{2}}{n_{th}+\frac{1}{2}}}\right)
\end{equation}
for an isotropic coherent state \cite{Carmichael1999,Walls2008}. In the
present quantum calculations, we address the zero-point fluctuation with $n_{th}%
\rightarrow0$. The probability distribution of the position and momentum
$P(\delta x_{0},\delta y_{0})$ (see e.g. Fig. \ref{fig3}(a) insets) is related
to the Wigner function as $P(\delta x_{0,\theta}^{\prime})=\int_{-\infty
}^{+\infty}W(\delta x_{0}^{\prime},\delta p_{0}^{\prime})d\delta p_{0}%
^{\prime}$, where $(\delta x_{0,\theta}^{\prime},\delta p_{0,\theta}^{\prime
})$ corresponds to $(\delta x_{0},\delta p_{0})$ rotated by $\theta$. For a
Gaussian, $P(\delta x_{0},\delta p_{0})=W(\delta x_{0},\delta p_{0})$. In
general, the Wigner function can be reconstructed from a measured $P$ by e.g.
a maximum likelihood or Radon transform \cite{Lvovsky2009}. The limit cycle of
$\pm\vec{k}$ modes in Fig. \ref{fig3}(a) is a circle with the same radius and
width as the corresponding Wigner function at the same $T_{\mathrm{env}}.$
$W(x_{0},p_{0})$ in Fig. \ref{fig4}(c) shows two local maxima pertaining to
two classically bistable points \cite{Drummond1980,Kheruntsyan1999}; the
self-Kerr bistability is a classical phenomenon, while quantum fluctuations
lead to finite distributions around the two fixed points in phase space.

\section{Squeezing of the noise}

\label{sec3} 
\begin{figure}[!]
\includegraphics[width=0.5\textwidth]{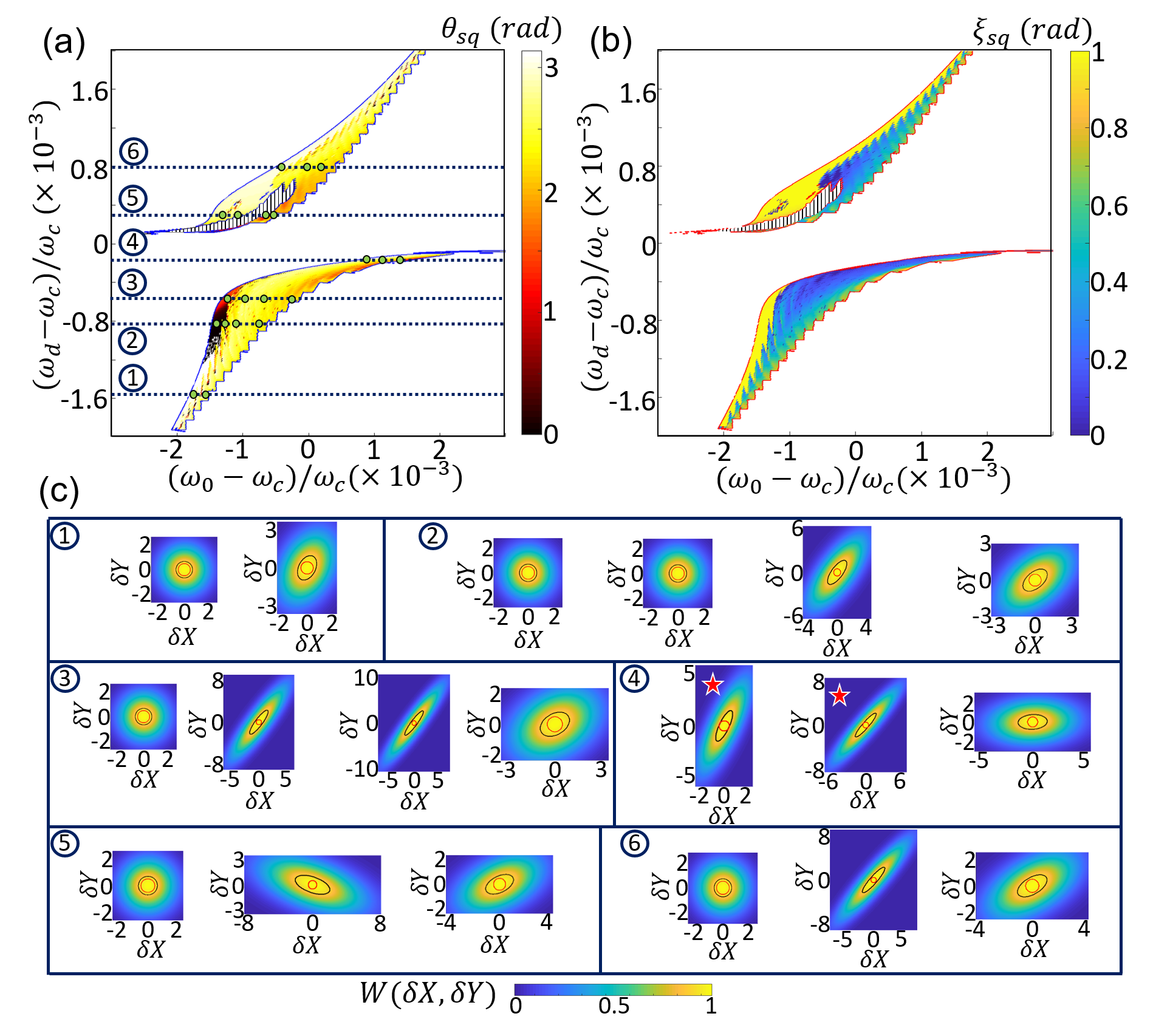}\caption{Fluctuation
ellipticity from steady state covariance matrix when $\vec{k}\neq0$ modes are
included. (a) Fluctuation variance ellipse angle $\theta_{sq}$. (b)
Fluctuation variance ellipticity $\xi_{sq}$. In (a) and (b), the hatched area
is where the classical steady state is a limit cycle or chaos (see Fig.
\ref{fig2}). (c) The constructed Wigner function of the fluctuation for the
cavity mode, $W(\delta X,\delta Y)$, from covariance matrix. Each numbered
section in (c) corresponds to $\omega_{d}$ indicated by a dashed line with the
same number, and $\omega_{0}$ of each panel is indicated by a hollow circle on
the corresponding dashed line (the left most panel of each section: case with
smallest $\omega_{0}$). The colormap is the same for all the panels, and
normalized to the max value of each panel.}
\label{fig5}%
\end{figure}

Since the quantum analysis rules out limit cycles in the steady
state of the Kittel (photon) mode, we may analyze their nature by focussing on
the phase space in the proximity of the fixed points FP1 and FP2, which is
accessible in terms of the steady state covariance matrix $\boldsymbol{\Lambda
}_{\infty}$ (see Appendix \ref{app2} for details), and experimentally in the cavity output field, e.g. by homodyne
detection \cite{Walls2008,Lvovsky2009}. In Figs. \ref{fig5}(a) and \ref{fig5}(b), we map
the angle of the minor axis $\theta_{sq}$ and ellipticity $\xi_{sq}$ of the
calculated variances. Figure \ref{fig5}(c) shows some examples of the
\textit{photonic} Wigner functions obtained from the covariance matrices
calculated by the quantum Langevin equations (non-scaled system with
$T_{\mathrm{env}}=1\,K$) \cite{Braunstein2005,Walls2008} as
\begin{align}
W(\delta X,\delta Y)  &  =\int_{-\infty}^{+\infty}d^{3}\delta x_{0(\pm\vec
{k})}d^{3}\delta p_{0(\pm\vec{k})}\frac{1}{(2\pi)\sqrt{\det\left(
\boldsymbol{\Lambda}_{\infty}\right)  }}\times\nonumber\\
&  \exp{\left\{  -\frac{1}{2}\boldsymbol{v\Lambda}_{\infty}v^{T}\right\}  },
\label{eq3}%
\end{align}
where $\boldsymbol{v}=[{\delta x_{0}},{\delta p_{0}},{\delta x}_{\vec{k}%
},{\delta p}_{\vec{k}},{\delta x}_{-\vec{k}},{\delta p}_{-\vec{k}},{\delta
X},{\delta Y}]$\textit{.} $d^{3}\delta x_{0(\pm\vec{k})}=d\delta x_{0}d\delta
x_{\vec{k}}d\delta x_{-\vec{k}}$, and $d^{3}\delta p_{0(\pm\vec{k})}=d\delta
p_{0}d\delta p_{\vec{k}}d\delta p_{-\vec{k}}$. Here, the black contours are the
computed variance ellipses, while the red circles indicate the zero-point
fluctuations of the non-interacting photon. This analysis only holds for fixed
points, otherwise the full Wigner function as in Figs. \ref{fig4}(b)-(f)
should be computed. For a classical state, the uncertainty in a given
direction in the position-momentum (quadrature) phase space can not be less
than 1\textit{, }which implies that the black and red contours may not
intersect. The two panels of Fig. \ref{fig5}(c) that are marked by purple
stars are therefore proof-of-principle that quantum squeezed states can be
generated. This magnon-induced squeezing of light is the first quantum
information resource reported here. It is an essential ingredient for accurate
measurements of e.g. sub-shot-noise phases in a Mach-Zehnder setup
\cite{Walls2008}. At certain points in the $(\omega_{0},\omega_{d})$ plane,
the amount of squeezing can be enhanced by increasing the power (i.e.
$\left\vert \alpha_{0,\pm\vec{k}}\right\vert $, see Appendix \ref{app2_1}) as
well as reducing $T_{\mathrm{env}}$.

\section{Entanglement}

\label{sec4} 
\begin{figure}[!]
\includegraphics[width=0.5\textwidth]{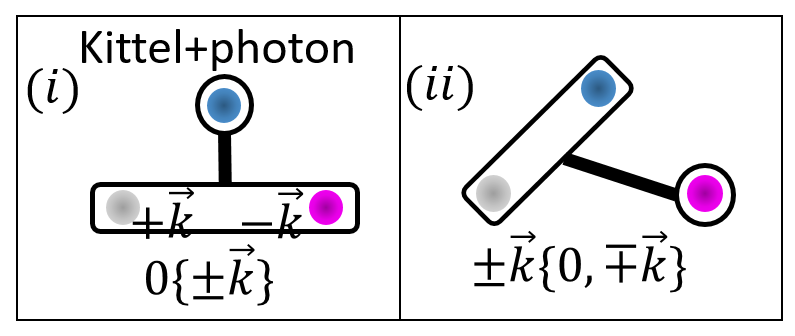}\caption{The two distinct
bipartite configurations of the essentially tripartite system, i.e. (i)
$0\{\pm\vec{k}\}$ and (ii) $\pm\vec{k}\{0,\mp\vec{k}\}$.}%
\label{fig6}%
\end{figure}

A second resource for quantum information is entanglement. The
4MS term of the form $c_{0}^{\dag}c_{0}c_{\vec{k}}^{\dag}c_{\vec{k}}%
+c_{0}^{\dag}c_{0}c_{-\vec{k}}^{\dag}c_{-\vec{k}}$ leads to a dominant
mean-field potential $\alpha_{0}\alpha_{\vec{k}}\delta c_{0}^{\dag}\delta
c_{\vec{k}}^{\dag}+\alpha_{0}\alpha_{-\vec{k}}\delta c_{0}^{\dag}\delta
c_{-\vec{k}}^{\dag}+\mathrm{h.c.,}$ i.e. a \textquotedblleft two-mode
squeezing\textquotedblright\ Hamiltonian for the Kittel and either mode of the
$\pm\vec{k}$ pair, which corresponds to the maximal bipartite entanglement of
two continuous variables \cite{Einstein1935,Braunstein2005}. When the
instability mixes Kittel with $\pm\vec{k}$ modes, i.e. $\alpha_{\pm\vec{k}%
}\neq0$, the modes should be entangled. Correspondingly, a 4MS term of the
form $c_{0}c_{0}c_{\vec{k}}^{\dag}c_{-\vec{k}}^{\dag}+c_{\vec{k}}^{\dag
}c_{\vec{k}}c_{-\vec{k}}^{\dag}c_{-\vec{k}}+\mathrm{h.c.}$ should entangle the
$\pm\vec{k}$ modes. In order to assess entanglement, we consider the
two distinct bipartite configurations as sketched in Fig. \ref{fig6}, (i) the
Kittel magnon-photon polariton as one part and the $\pm\vec{k}$ pair as the other,
and (ii) one of the modes of the $\pm\vec{k}$ pair, say $\vec{k},$ considered
as one part and $-\vec{k}$ plus the Kittel magnon-photon polariton as the other. It should be reminded that the Kittel mode and the cavity photon form a hybridized mode, polariton, due to the strong coupling through the beam splitter interaction $ic_0b^\dag+h.c.$, and can be considered as one part. For
applications such as quantum teleportation \cite{Braunstein1998,Furusawa1998}, the
entanglement of distillation $E_{D}$, i.e. the rate of entanglement (number of
perfect Einstein-Podolsky-Rosen states \cite{Einstein1935} such as spin
singlets) is an important parameter. It can be extracted from a bipartite
state by local operations and classical communication (the so-called LOCC
protocols) \cite{Braunstein2005,Bennett1996,Vidal1999,Vidal2000}. Here we
employ the concept of negative partial transposition (NPT)
\cite{Peres1996,Horodecki1996}: the existence of negative eigenvalues of a
bipartite state density matrix $\rho_{1}\otimes\rho_{2}$ after partial
transposition $(\rho_{1})^{T}\otimes\rho_{2}$ is a sufficient condition for an
entangled state, for bipartite Gaussian states even a necessary one
\cite{Braunstein2005}. The degree of negativity in the partial transposed
density matrix can be quantified by the logarithmic negativity $E_{LN}$ which
determines the upper bound of $E_{D}$ \cite{Vidal2002,Braunstein2005}. In
other words, depending on the LOCC protocols used for purification of
entanglement of a general mixed state of a bipartite configuration, a maximum
number of $E_{LN}\times N$ entanglement bits (number of spin-singlets) can be
distilled, where $N$ is the number of copies of the bipartite state. We are
therefore interested in the logarithmic negativity $E_{LN}$
\cite{Peres1996,Horodecki1996,Bennett1996,Vidal2002,Braunstein2005} of both
our semi-classical and quantum density matrices. The former is calculated from
the covariance matrix and the latter directly from the density matrix in the
number space of involved modes (see Appendix \ref{app3}). Another measure is
the entanglement of formation $E_{F}$, i.e. the number of fully entangled
bipartite particles (such as a spin-singlet) required to form the state, or in other words the
minimum of the von Neumann entropy of the bipartite state among different
(infinite) realizations of a mixed state. We compute $E_{F}$ for completeness,
but note that in contrast to $E_{D},$ its value is not of practical
importance. The details of the calculations for both approaches to the density
matrix are deferred to Appendix \ref{app4}, which also provides a short review
of the entanglement measures used here.

\begin{figure}[!]
\includegraphics[width=0.5\textwidth]{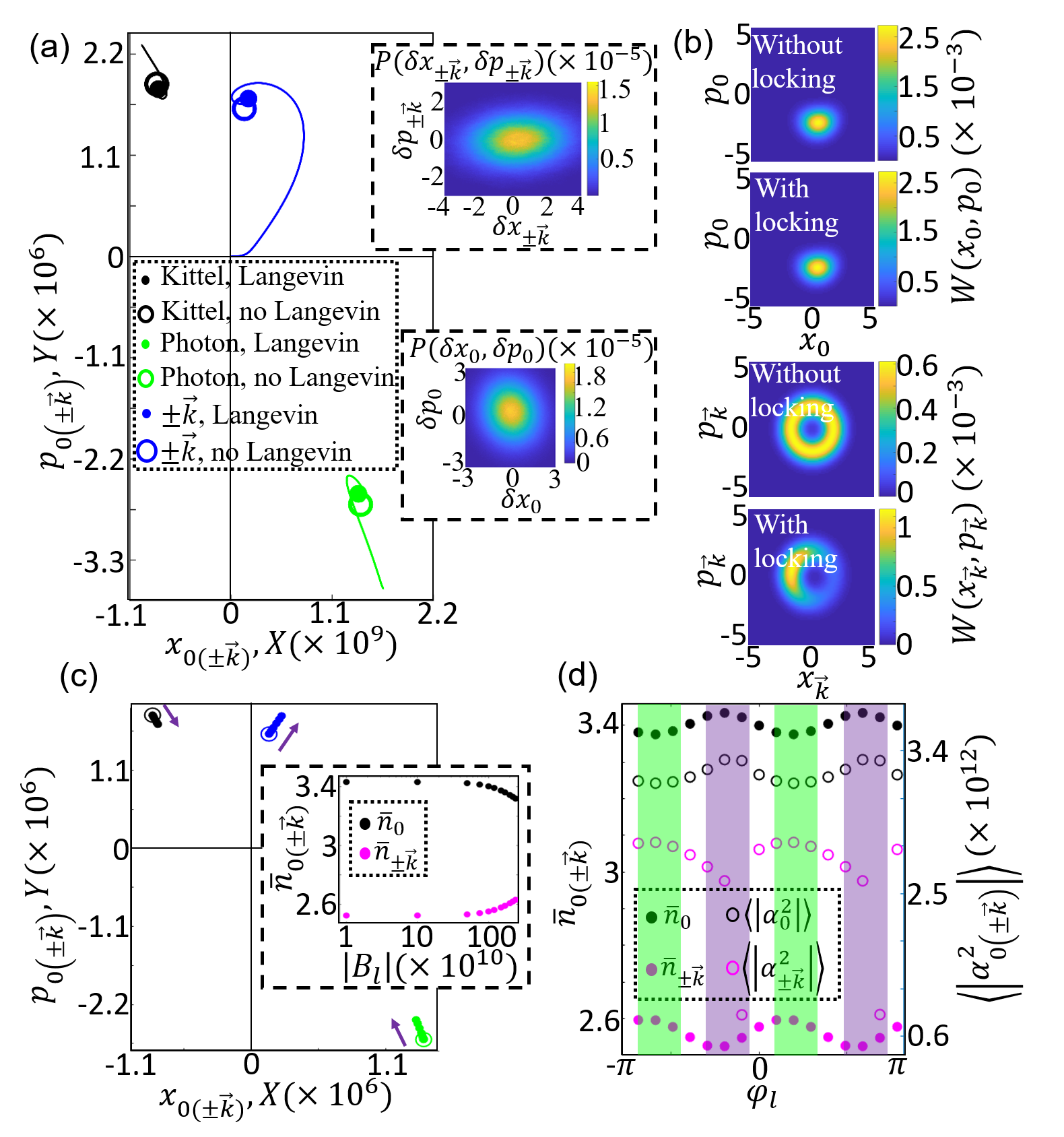}\caption{Injection locking
of $\pm\vec{k}$ modes calculated by (a) the Langevin formalism and (b) the
quantum master equation at the $(\omega_{0},\omega_{d})$ of the black star in
Fig. \ref{fig2}. (a) Main panel: The averaged trajectories from the initial to
the final states (filled dots) for all modes. $\left\vert B_{l}\right\vert
=10^{12}\,1/\mathrm{s}\left(  <\bar{B}/10\right).$ The
color code\ explained in the panel is applicable also to (c). Insets:
Probability distributions of the fluctuations. The final states from entirely
classical calculations and without injection locking are also shown. (b)
Results from the quantum master equation, without and with injection locking
as indicated. (c) Main panel (Langevin): the averaged final states of all
fields as a function of locking field $B_{l}$ from $10^{10}$ to $2\times
10^{12}\mathrm{\,s}^{-1}$ with fixed $\phi_{l}=0.$ The purple vectors indicate
the shift direction with increasing $B_{l}$. Inset: Mean magnon number
$\bar{n}_{0(\pm\vec{k})}$ in the scaled quantum system as a function of
$B_{l}$ with $\phi_{l}=0$. (d) Dependence of the averaged $|\alpha_{0(\pm
\vec{k})}|^{2}$ for the final states from Langevin (right axis) and $\bar
{n}_{0{\pm\vec{k}}}$ from quantum master equation calculations (left axis), on
the phase $\phi_{l}\in\left\{  -\pi,\pi\right\}  $ for a fixed $|B_{l}%
|=10^{12}\,\mathrm{s}^{-1}$. The purple (green) colored background emphasizes
$\phi_{l}$ values with weak (strong) locking.}%
\label{fig7}%
\end{figure}

Unfortunately, we find that the covariance matrix extracted from the Langevin
formalism leads to $E_{LN,0\{\pm\vec{k}\}(LN,\pm\vec{k}\{0,\mp\vec{k}\})}%
^{L}=0$ (superscript $L$ stands for Langevin) all over the $(\omega_{0}%
,\omega_{d})$ map: The trace over one mode of an imperfect two-mode squeezed
state leads to a (broadened) squeezed coherent state with an almost deterministic
phase. Moreover, the relative position and total momentum of a two-mode
squeezed state is also (almost) deterministic. For example, when the bipartite
state of the Kittel mode and, say, $+\vec{k}$ mode of the $\pm\vec{k}$ pair is
close to a two-mode squeezed state, $x_{0}-x_{+\vec{k}}$ and $p_{0}%
+p_{+\vec{k}}$ should be deterministic. However, since the attractors of the
$\pm\vec{k}$ modes are limit cycles [see Fig. \ref{fig3}(a) and Figs.
\ref{fig4}(c)-(e)], while that of the Kittel mode is a fixed point,
$x_{0}-x_{+\vec{k}}$ and $p_{0}+p_{+\vec{k}}$ are undetermined, no distillable
continuous variable entanglement should be expected. The analysis based on the
covariance matrix is only accurate for Gaussian states or continuous variable
entanglement \cite{Braunstein2005}, so $E_{LN}^{L}=0$ does not contradict the
finite distillable bipartite entanglement found in the quantum solutions of
the scaled system, which indicate that the states are non-Gaussian. For
example, at $(\omega_{0},\omega_{d})$ indicated by the black star in Fig.
\ref{fig2} (all the results in Fig. \ref{fig7} and Fig. \ref{fig8} correspond
to this point), $E_{LN,0\{\pm\vec{k}\}}^{q}\sim0.3$ and $E_{LN,\pm\vec
{k}\{0,\mp\vec{k}\}}^{q}\sim0.4$ (superscript $q$ stands for quantum), whereas
the continuous variable entanglements $E_{LN,0\{\pm\vec{k}\}(LN,\pm\vec
{k}\{0,\mp\vec{k}\})}^{L}=0$. In other words, entangled states of continuous
variables do not comprise all of the entanglement conceivable from bosonic
modes. For two modes, a state such as $|0,N\rangle+|N,0\rangle$ (0 is the
vacuum Fock state in one mode and $N$ is the $N$'th Fock state in the other
mode) is a maximally entangled state \cite{Fiurasek2002,*Lee2002} but not a
two-mode squeezed state. 

\section{Distillable Gaussian entanglement}

\label{sec5} The finite distillable entanglement by non-Gaussian states
predicted for the driven magnet can be assessed experimentally only by a full
reconstruction of the density matrix, which is technically very challenging.
Only very recently techniques for quantum state tomography of Gaussian states
in microwave frequencies have been developed (see below). On the other hand,
Gaussian states are fully characterized by the second moment or the auto and
cross-correlations which are more readily measured and sufficient to assess
the continuous variable entanglement. However, this requires getting rid of
the limit cycles. This can be achieved by fixing the phases of the $\pm\vec
{k}$ modes via \textquotedblleft injection locking\textquotedblright%
\ \cite{Slavin2009,Lee2013,Aspelmeyer2014,Demidov2014} by an AC coherent drive
with fixed phase, a standard technique used e.g. to improve current-induced
spin oscillations \cite{Demidov2014}.

\subsection{Injection locking of $\pm\vec{k}\neq0$ magnons}

\label{sec5_1} Here we study a spatially modulated injection locking with
Hamiltonian $H_{l}=i(B_{l}e^{-i\omega_{L}t}c_{\pm\vec{k}}^{\dag}-B_{l}^{\ast
}e^{i\omega_{L}t}c_{\pm\vec{k}})$ with drive frequency $\omega_{L}=\omega
_{\pm\vec{k}}$ and amplitude $B_{l}=\left\vert B_{l}\right\vert \exp(i\phi
_{l})$ and phase $\phi_{l}$, which couples to both modes of the $\pm\vec{k}$
pair. Large enough $B_{l}$ transforms limit cycles into fixed points, in both
the semi-classical [see Fig. \ref{fig7}(a)] and quantum calculations [see Fig.
\ref{fig7}(b)]. Figure \ref{fig7}(c) illustrates the effect of locking as a
function of $|B_{l}|$ and a fixed phase $\phi_{l}=0$. With increasing
$\left\vert B_{l}\right\vert $, the mean number of $\pm\vec{k}$ magnons
increases, whereas the numbers of Kittel mode magnons and photons decrease.
The effects are small but establish identical trends in both Langevin and
quantum formalisms. According to Fig. \ref{fig7}(d), the phase $\phi_{l}$
modulates the excitations with periodicity of $\pi$, since the force is proportional to
$\cos(\phi_{l}+\phi_{0}),$ where $\phi_{0}$ is a constant shift
\cite{Slavin2009}. Strong (weak) locking implies larger (smaller) number of
$\pm\vec{k}$ magnons, and smaller (larger) number of Kittel mode magnons and
photons both in Fig. \ref{fig7}(c) and (d): The mean magnon numbers $\bar
{n}_{0(\pm\vec{k})}$ in the scaled quantum system follow the trends of the
equivalent $|\alpha_{0(\pm\vec{k})}|^{2}$ of the Langevin formalism.

\subsection{Injection locking and Gaussian distillable entanglement}

\label{sec5_2} 
\begin{figure}[!]
\includegraphics[width=0.5\textwidth]{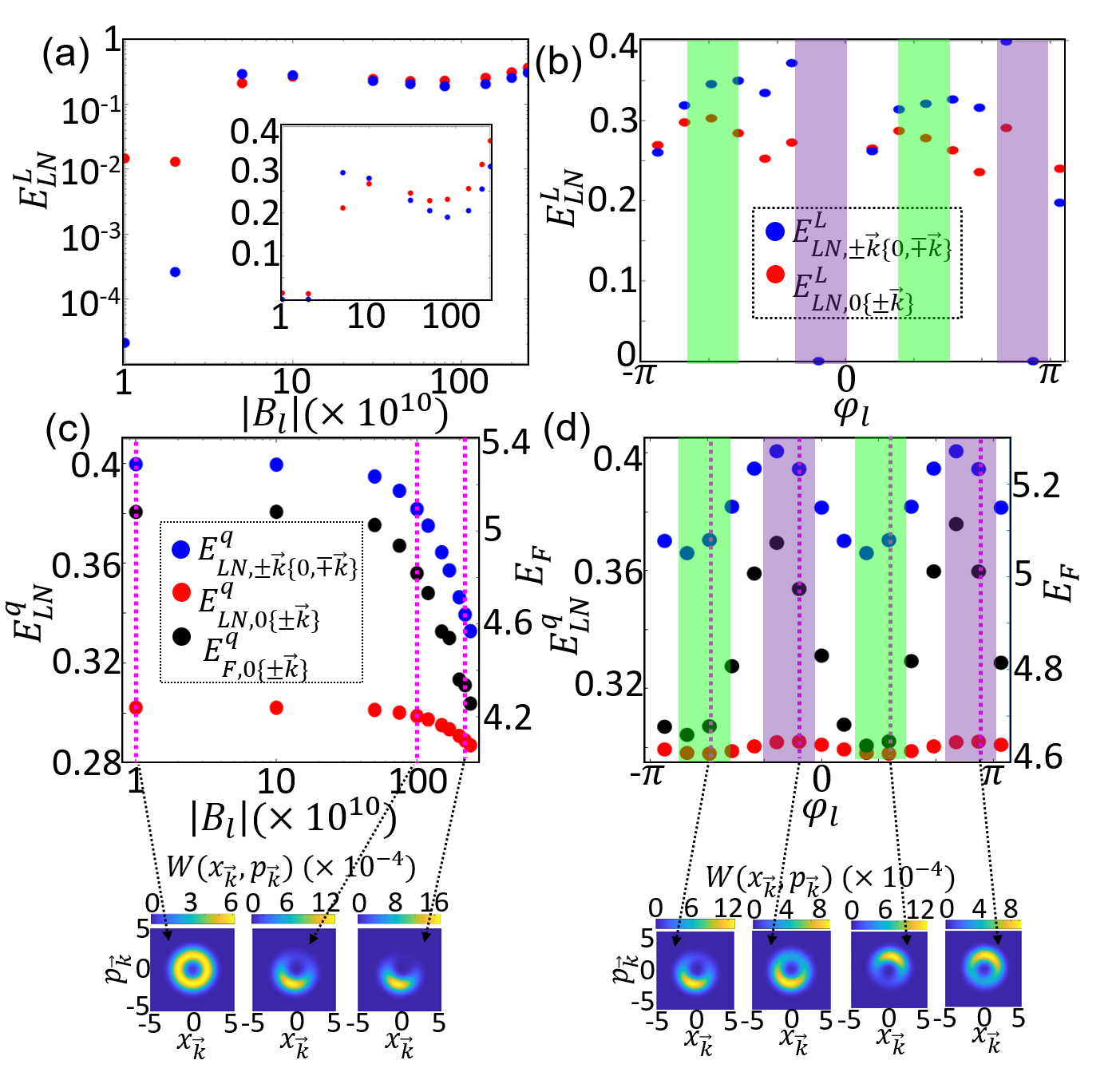}\caption{Entanglement in an
excited magnon system with injection locking. (a) The logarithmic negativity
$E_{LN}^{L}$ of steady states from the Langevin formalism as a function of
$|B_{l}|$ for $\phi_{l}=0$.\ In the main panel, both axes are on a log scale.
The inset contains the same data on a linear scale. (b) $E_{LN}^{L}$ as
function of $\phi_{l}$, at $\left\vert B_{l}\right\vert =10^{12}%
\mathrm{s}^{-1}$.\textit{ }(c) and (d) Logarithmic negativities (left axis)
$E_{LN,0{\pm\vec{k}}}^{q}$, $E_{LN,\pm\vec{k}{,0\mp\vec{k}}}^{q}$, and
entanglement of formation (right axis) $E_{F,0{\pm\vec{k}}}$, in the steady
state, calculated by the quantum master equation as a function of $\left\vert
B_{l}\right\vert $ for $\phi_{l}=0$ and as a function of $\phi_{l}$ for
$\left\vert B_{l}\right\vert =10^{12}\mathrm{s}^{-1}$, respectively. Bottom
insets of (c) and (d) are Wigner functions of $\pm\vec{k}$ modes for
particular values of $\left\vert B_{l}\right\vert $ and $\phi_{l}$,
respectively, indicated by black dashed arrows and purple dashed lines. In (b)
and (d), the purple (green) colored bars indicate $\phi_{l}$ values
corresponding to weak (strong) locking. }%
\label{fig8}%
\end{figure}

The beneficial effects of locking on the logarithmic negativity
extracted from the semi-classical covariance matrix for continuous variable
entanglement as a function of $\left\vert B_{l}\right\vert $ and $\phi_{l}$
are evident in Figs. \ref{fig8}(a) and \ref{fig8}(b). Figure \ref{fig8}(a) shows that
both $E_{LN,0\{\pm\vec{k}\}}^{L}$ and $E_{LN,\pm\vec{k}\{0,\mp\vec{k}\}}^{L}$
become nonzero by increasing $\left\vert B_{l}\right\vert $ to values where
locking is achieved and reach $\sim0.3$, which is in the range of the
predicted $E_{LN}^{q}$ without injection locking. Therefore, the distillable
entanglement became that of continuous variables and is accessible from the
covariance matrix. Figure \ref{fig8}(b) shows that $E_{LN,0\{\pm\vec{k}%
\}(\pm\vec{k}\{0,\mp\vec{k}\})}^{L}$ strongly depends on the phase $\phi_{l}$.
For the $\phi_{l}$ with weakest locking, $E_{LN}^{L}=0$.

\subsection{Effect of injection locking on entanglement}

\label{sec5_3} As mentioned earlier, in contrast to the general one,
continuous-variable bipartite entanglement requires injection locking. We
assess the former by studying the density matrix of the scaled system that solves the
quantum master equation, using the logarithmic negativity and entanglement of
formation $E_{F}$. The entanglement of formation $E_{F}$
\cite{Bennett1996_1,Wootters2001} in the bipartite configuration (i) is
calculated by the algorithm \cite{Zyczkowski1999} explained in Appendix
\ref{app4_2}. It should be noted that for a mixed state $E_{F}$ can be very
different from (but always larger than) the distillable entanglement
\cite{Wootters2001,Adesso2005}. Figures \ref{fig8}(c) and (d) show the
dependence of $E_{LN,0\{\pm\vec{k}\}}^{q}$, $E_{LN,\pm\vec{k}\{0,\mp\vec{k}%
\}}^{q}$, and $E_{F,0\{\pm\vec{k}\}}$ on locking field amplitude $\left\vert
B_{l}\right\vert $ and phase $\phi_{l}$. All entanglement measures are nonzero
without the injection locking, and remain finite when locking is added.
However, a stronger locking somewhat reduces $E_{LN,0\{\pm\vec{k}\}}^{q}$,
$E_{LN,\pm\vec{k}\{0,\mp\vec{k}\}}^{q}$, and $E_{F}$, in contrast to
$E_{LN,0\{\pm\vec{k}\}}^{L}$, $E_{LN,\pm\vec{k}\{0,\mp\vec{k}\}}^{L},$ which
are strongly enhanced by it. At large locking amplitudes $E_{LN}^{q}\approx
E_{LN}^{L},$ as expected for Gaussian states. The colored background in Figs.
\ref{fig8}(b) and (d) codes the regions with stronger (green) and weaker
(purple) effects of locking. In particular, Wigner functions of the $\pm\vec{k}$
modes in Figs. \ref{fig8}(c) and (d), display more ring-like (coherent state)
features, which explain weaker (stronger) effects of locking. Phase-locking
the existing $\pm\vec{k}$ magnons induces a fraction of magnons on top of
those generated by the instability of the Kittel mode, since they are driven
by both $i(B_{l}e^{-i\omega_{L}t}c_{\pm\vec{k}}^{\dag}-B_{l}^{\ast}%
e^{i\omega_{L}t}c_{\pm\vec{k}})$ and the (mean-field) 4MS term $\mathcal{D}%
^{4MS,2}_{0,\vec{k}}\alpha_{0}^{2}\alpha_{\pm\vec{k}}c_{\mp\vec{k}%
}+\mathrm{h.c.}$ (see Appendix \ref{app1}). $\mathcal{D}^{4MS,2}_{0,\vec{k}%
}\alpha_{0}^{2}\alpha_{\pm\vec{k}}\sim10^{12}\,\mathrm{s}^{-1}$ has the same order of magnitude as $B_{l}^{\prime}$. Stronger locking reduces the number of
magnons paired with the Kittel mode magnons (i.e. generated from 4MS terms),
and therefore the entanglement measures $E_{LN}^{q}$ and $E_{F}$ as observed
in Figs. \ref{fig8}(c) and \ref{fig8}(d). We compare the dependence of
$E_{LN}^{L}$ and $E_{LN}^{q}$ on $\phi_{l}$ in Figs. \ref{fig8}(b) and \ref{fig8}(d).
$E_{LN}^{L}$ ($E_{LN}^{q}$) is larger (smaller) for $\phi_{l}$ corresponding
to stronger (weaker) locking, even vanishes at some phase angles for which the
limit cycle is not significantly suppressed. The inequality $E_{F}\geq E_{LN}$
is always obeyed.

\section{Experimental relevance}

\label{sec6} 
\begin{figure}[!]
\includegraphics[width=0.5\textwidth]{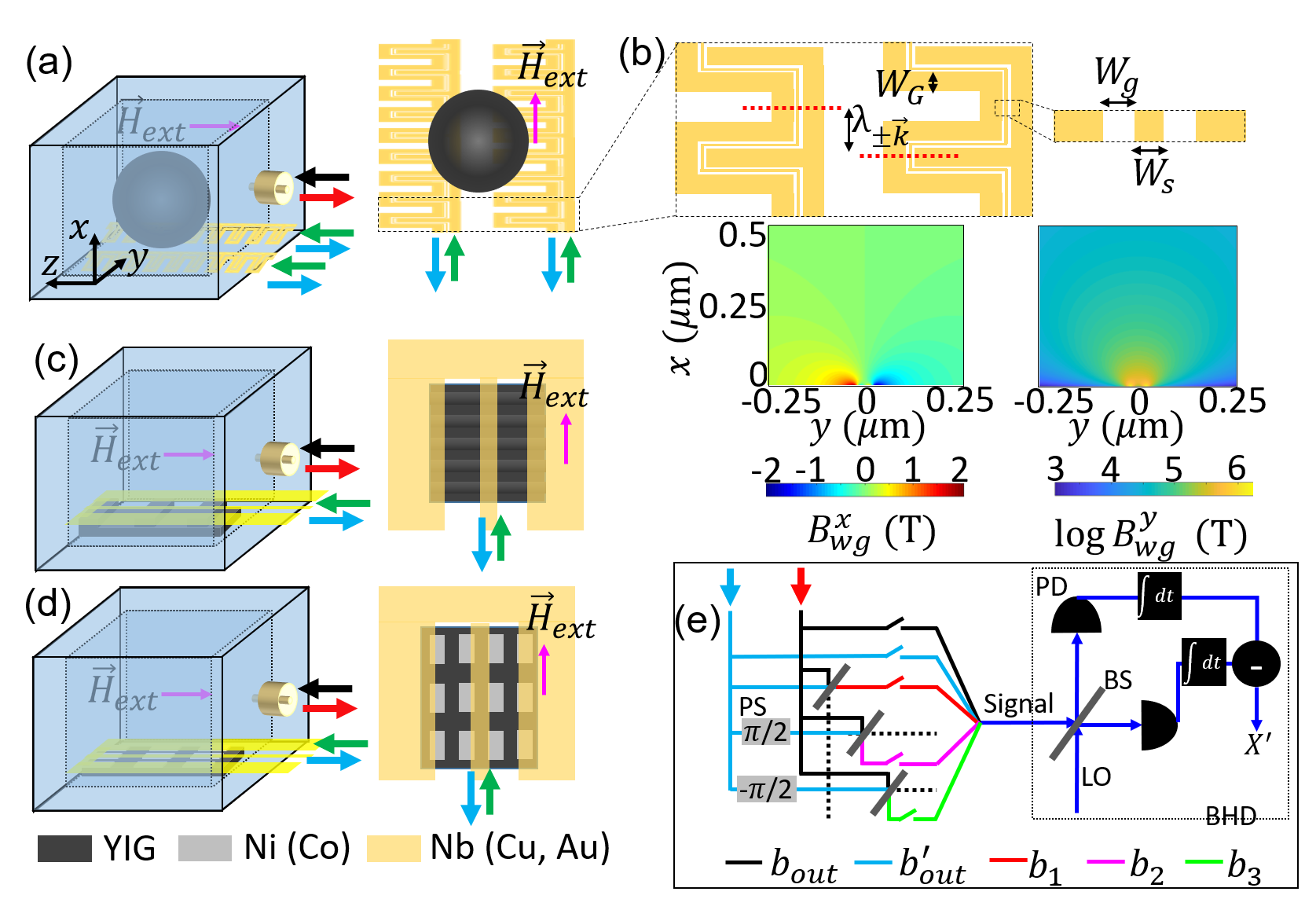}\caption{Proposed set-ups to
measure entanglement in magnets. (a), (c) and (d) A magnetic sphere or film in
a rectangular cavity with wave guides for injection locking. The black and red
arrows indicate the input drive $\bar{B}$ and output microwave fields
($b_{out}$), respectively. The green arrows are the locking field $B_{l}$
applied by a coplanar waveguide and the blue arrow is the corresponding output
signal ($b_{out}^{\prime}$). The suggested materials for (a), (c) and (d) are
indicated at the bottom of panel (d). (b) Top panel: Zoom-in of a section of
the staggered waveguide in (a), and the dimensions $W_{G}$, $\lambda_{\pm
\vec{k}}$, $W_{g}$, and $W_{s}$. Bottom panel: Magnetic field distribution
along either of the red dashed lines in the top panel, for $W_{s}%
=W_{g}=10\,\text{nm}$ and voltage $V_{wg}=10^{-2}\,\text{V}$. (e) Equivalent
electronic circuit model including beam splitters (BS), phase ($\pm\pi/2$)
shifters (PS), switches, photon number detectors (PD), temporal integrators
(indicated by $\protect\int dt$), a subtractor [indicated by ($-$)], signal
and local oscillator (LO) lines. The color code of the signal lines is given
in the bottom of the panel. The dashed black lines are unused signals. The
output $X^{\prime}$ is governed by the LO-phase, while the signal line forms
the input to the homodyne detection (the parts enclosed by the dashed black
rectangle).}%
\label{fig9}%
\end{figure}

The injection locking Hamiltonian and distillation of the
entanglement requires coupling of an AC magnetic field to the $\pm\vec{k}$
magnons, i.e., a spatial modulation in the dynamic magnetization with period
$\sim\lambda_{\pm\vec{k}}=2\pi/k\sim100\,$nm, which can be accomplished by
gratings such as a periodically modulated coplanar waveguide close to either
sphere or thin film [Figs. \ref{fig9}(a) and \ref{fig9}(b)], periodic trenches in a YIG
thin film [Fig. \ref{fig9}(c)] \cite{Chumak2014}, or a 2D ferromagnetic
nanowire array deposited on top of a thin film [Fig. \ref{fig9}(d)]
\cite{Chen2018}. In Fig. \ref{fig9}(d), nano-structured magnets act as
in-phase antenna for the microwave input at frequency $\omega_{\pm\vec{k}}$
\cite{Chen2018,*Yu2019_1,*Chen2019}. For a sphere, a periodic waveguide [Figs. \ref{fig9}(a) and
\ref{fig9}(b)] appears to be the only viable method, but microwave fields lose their
modulation with distance. The quality of the spatial modulation improves with
reduced size of the sphere to say tens of $\mathrm{\mu}\text{m}$. The $k$
value of interest is not affected by the size of the magnetic element down to
a radius of $>1\,\mu\text{m}$, and therefore the periods in the proposed
structures in Figs. \ref{fig9}(a)-(d) do not have to be scaled. The parameters
of the staggered waveguide in Figs. \ref{fig9}(a) and (b) are the signal wire
width $W_{s}$ and the width of the gap between signal and ground lines $W_{g}%
$. Figure \ref{fig9}(b) also shows a snapshot of the AC magnetic field
$\vec{B}_{wg}=B_{wg}^{x}\hat{x}+B_{wg}^{y}\hat{y}$ \cite{Carlsson1999}, for
$W_{s}=W_{g}=10\,\text{nm}$ and input voltage $V_{wg}=10\,\mathrm{m}\text{V,
}$which governs $B_{l}$ as a function of the waveguide input power $P_{in,wg}%
$. It is periodic in $z$ with wave length $\lambda_{\pm\vec{k}}$.
Here $B_{wg}^{x}$ may be disregarded because the integral of $B_{wg}^{x}%
m_{\pm\vec{k}}^{x}$ vanishes, where $m_{\pm\vec{k}}^{x}\left(x,y\right)  $
is a magnetization of a volume mode with $\vec{k}\|\hat{z}$. On the other hand, $B_{wg}^{y}$ has finite overlap with ${m}_{\pm\vec{k}}^{y}\left(x,y\right)$ since modulated by the same
wavelength. We can quench an unwanted coupling to the Kittel mode by a $\pi/2$
phase shift of the input power between the two wave guides. For a cube with
$V_{m}=\left(100\,\mathrm{\mu}\text{m}\right) ^{3}$, we chose $\bar{B}%
\sim10^{13}\,\text{s}^{-1}$ and $B_{l}=\gamma\sqrt{M_{s}%
V_{m}/\gamma\hbar}B_{wg}^{\prime y}=10^{12}\,\mathrm{s}^{-1}$ where average
$B_{wg}^{y}$ over the magnet $B_{wg}^{\prime y}\sim0.2\,\mathrm{mT}$. Based
on the field distribution $B_{wg}^{y}$ in Fig. \ref{fig9}(b) which is for
$V_{wg}=10\,$mV, a simple approximation shows that the required $B_{wg}%
^{\prime y}$ is thus obtained using $V_{wg}=\sqrt{2P_{in,wg}Z_{0}}=0.1\,$V or a
waveguide input drive power $P_{in,wg}\sim0.1\,$mW and impedance $Z_{0}%
\sim50\,\Omega$. For a $10\,\mu$m cube, $V_{wg}=\sqrt{2P_{in,wg}Z_{0}}=1\,$mV or a
waveguide input drive power $P_{in,wg}\sim10\,$nW is adequate to provide the required $B_{l}\sim10^{11}\,$1/s. This is a small perturbation on top of the cavity drive,
which for $V_{m}=\left(10\,\mathrm{\mu}\text{m}\right)^{3}$ is
$P_{in}=10\,\mathrm{\mu}\text{W}$ corresponding to $\bar{B}\sim 10^{12}\,$1/s.

The microwaves that drive magnons out of equilibrium, heat the magnetic system by
Gilbert damping to temperatures above the assumed $T_{\mathrm{env}}\sim1\,$K,
so we have to assess the conditions at which our theory remains
applicable. Let us assume a lattice temperature of the magnet of
$T_{\mathrm{L}}$ and an environmental one $T_{\mathrm{env}}$. We assume that the
magnet is in contact with an acousticlly matched material such as gadolinium
gallium garnet (GGG) with temperature equal to $T_{\mathrm{L}}$ at the contact
to YIG and $T_{\mathrm{S}}$ at the other side, which can be kept lower than $T_{\mathrm{env}}$ if actively cooled. Geometrically, we assume a sample which is narrow and suddenly becomes wide. The wide part of the sample is the magnetic cube kept at $T_L$ and there is a gradient in the narrow one, GGG substrate. In the steady state, the power input
$P_G$ by Gilbert damping must equal the heat current through the substrate
$P_G=G_{\mathrm{GGG}}(T_{\mathrm{L}}-T_{\mathrm{S}})$, where the heat
conductance $G_{\mathrm{GGG}}=\sigma_{\mathrm{GGG}}d_{\mathrm{GGG}},$ with
thermal conductivity $\sigma_{\mathrm{GGG}}\sim7\,\text{W}/\left(
\text{K}\text{m}\right)  $ \cite{Langenberg2016} and $d_{\mathrm{GGG}}$ is the
thickness of the substrate. $P_G=\hbar\omega_{0}|\alpha_{0}|^{2}\zeta_{m,0}$,
where $\zeta_{m,0}$ is the Kittel magnon dissipation rate \cite{Brataas2011}. A magnon
number $|\alpha_{0}|^{2}$ less than $10^{14}$ [see e.g. Fig. \ref{fig1}(c)]
causes $P_G\sim10^{-3}\,\text{W}$. Therefore the heat sink temperature should be
kept at $T_{\mathrm{S}}\sim T_{\mathrm{L}}-\left(  d_{\mathrm{GGG}%
}/1\,\mathrm{mm}\right)  \,$K. For a smaller magnet with $10\,\mathrm{\mu
}$m dimension, $P_G\sim10^{-6}\,\text{W}$ and for $d_{\mathrm{GGG}}\sim1\,\mu$m,
$T_{\mathrm{S}}\sim0.9\,$K is adequate to keep $T_{\mathrm{L}}\sim1\,$K. A magnetic sphere attached to the
cooling system by a glue may face a higher heat resistance to the heat sink. The same statement applies to the resistance of the interface of the heat sink and the environment. The relevant parameter is then the total interface heat conductance of the magnet with environment. The worst case scenario corresponds to the magnetic sphere (cube) free standing in the environment (no heat sink), and $P/S_{M}=G_{LE}(T_{L}-T_{\mathrm{env}})$ is applicable, where $S_{M}$ is the surface of the magnet, and $G_{LE}$ is the effective interface thermal conductance of magnet lattice/environment interface.
When $T_{\mathrm{env}}$ is in tens of mK regime and magnet dimension is $10\,\mu$m,
$G_{LE}\sim10^{4}\text{W}/\text{K}\text{m}^{2}$ is adequate to keep $T_{L}\sim 1\,$K, which can
be achieved in cryogenic environments \cite{Smith1969,*Baudouy2015}. In presence of heat sink, conditions are more relaxed, i.e. smaller (larger) $G_{LE}$ ($T_{\mathrm{env}}$) can still lead to desired $T_{L}\sim 1\,$K.

Finally, we address the observability of the predicted entanglement in the
set-ups of Figure \ref{fig9}. The input drive to the cavity and the waveguide
should be locked by a tunable phase shifter for the waveguide input. The
output cavity field contains the information on the amplitude and squeezing of
the Kittel mode (see Fig. \ref{fig5}). The waveguide output reveals essential
information about the statistics of the $\pm\vec{k}$ pairs, e.g. the
limit-cycle attractors [see Fig. \ref{fig3}(a)] and in the case of
injection-locking, fixed-point dynamics (see Fig. \ref{fig7}). Both outputs
are required to detect bipartite entanglement of the Kittel mode and the
$\pm\vec{k}$ pair. Figure \ref{fig9}(d) illustrates the homodyne tomography
\cite{Lvovsky2009} method implemented in \cite{Wenger2005,Dauria2009}. The
balanced homodyne detection (BHD) output $X^{\prime}$ determines the
quadrature statistics of the input signal by varying the phase of the local
oscillator (LO) and measuring $X^{\prime}$ many times to reveal its first and
second moments \cite{Lvovsky2009,Walls2008}. For example, the BHD detection\ of the cavity field
characterizes $X^{^{\prime}}$ as a function of LO phase, which leads to the
Wigner functions of Fig. \ref{fig5}(c). For bipartite entanglement, the
corresponding covariance matrix should be evaluated, which consists of two
diagonal and two off-diagonal blocks. The former are evaluated by feeding
either the output of the cavity $b_{out}$ or the output of the waveguide
$b_{out}^{\prime}$ to the BHD. For the off-diagonal blocks, the feed should be
$b_{1}=(b_{out}+ib_{out}^{\prime})/\sqrt{2}$, $b_{2}=(b_{out}-b_{out}^{\prime
})/\sqrt{2}$, and $b_{3}=(b_{out}+b_{out}^{\prime})/\sqrt{2}$
\cite{Dauria2009}, while both quadrature statistics should be evaluated. The
fields $b_{1}$, $b_{2}$, and $b_{3}$ can be filtered out by phase shifters and
mixing at beamsplitters, as depicted in Fig. \ref{fig9}(e). From the
covariance matrix elements, $E_{LN,0{\pm\vec{k}}}$ can be extracted.

The sketched quantum state tomography is a mainstream technique for quantum
information studies with light. The large photon energy renders single photon
detection relatively easy. Similar experiments in the microwave regime have
been carried out only recently \cite{Menzel2010,Bozyigit2011,Menzel2012}.
Standard microwave components such as high electron mobility transistors and
linear detectors, as well as proper design of beam splitters
\cite{Mariantoni2010,Hoffmann2010} led to development of techniques suitable
for microwave quantum state tomography. These techniques demonstrated and
characterized path entanglement \cite{Menzel2012}, as is required to test our predictions.

\section{Conclusion}

Quantum information and its resources such as squeezing and entanglement have
been pursued for discrete variables \cite{Makhlin2001}, continuous variables
in position-momentum phase space \cite{Braunstein2005}, and continuous
variables on the Bloch sphere phase space \cite{Arecchi1972}. Both discrete
and continuous variable systems have been considered for quantum computation
\cite{DiVincenzo1995,Lloyd1999}, while squeezed and entangled photons or
magnons \cite{Braunstein2005,Kitagawa1993,*Esteve2008,*Riedel2010} displayed
sub-shot noise and quantum teleportation
\cite{Giovannetti2004,Vaidman1994,*Furusawa1998}. Non-classicalities in
continuous variables \cite{Qian2012,Vlastakis2013,McConnell2015} pave the way
for future quantum computation protocols \cite{Veitch2013}. Therefore,
continuous variable quantum information offers more opportunities than that
with discrete variables, simply because the accessible Hilbert space is
larger. Here, we uncovered the continuous variable quantum information
resources of a coherently driven magnet, promising a large steady-state
deterministic entanglement by virtue of the nonlinearities and anisotropies of
the magnetization dynamics. For that reason we believe that magnons can beat
alternative systems such as mechanical membranes and cantilevers coupled to
cavity photons \cite{Aspelmeyer2014} as squeezing and entanglement resources.
The GHz magnons in our scheme are strongly entangled in magnets down to the
micrometer regimes, which is diffult to achieve in mechanical systems. The
scalability could help profit from the flexibility of artificial
metamaterials, such as arrays of $N$ nanomagnets on top of a waveguide [see
e.g. Fig. \ref{fig9}(a)], thereby accessing a large amount of deterministic
bipartite entangled states ($\sim0.3\times N$). More theoretical and
experimental efforts to utilize magnets in quantum information are needed,
however. The viscous damping of the magnetization dynamics is larger than that of
other systems. While this is of little direct concern for coherently driven
systems, the associated temperature increase must be controlled by advanced
heat management. Nanostructuring of high quality magnets is required, but
still in its infancy \cite{Zhu2017,Heyroth2019}.

Another direction to be pursued is based on treating the magnetostatic
manifold as atomic levels and assess quantum information resources on Bloch
spheres corresponding to two such levels. The latter can provide a link to
laser induced spin-orbit coupling \cite{Liu2009,Wu2016} of magnons and
topology in a single magnet. Moreover, the inherent chirality of magnon-photon
coupling \cite{Tao2019} can be employed to achieve light induced spin-orbit
coupling of magnons in more than one magnet and magnonic lattices with
controllable topology \cite{Wu2016}. These would allow topological quantum
information protocols \cite{Nayak2008} or unique features such as robustness
of edge states, and conventional quantum information as addressed in this
paper, coexisting in a single magnet or a collection of them \cite{Peano2016}.

\vskip0.25cm \begin{acknowledgments}
This work is financially supported by JSPS KAKENHI (Grant No. 19H006450) and the Nederlandse Organisatie voor Wetenschappelijk Onderzoek (NWO). M. E. conducted part of this research at Kavli Institute of Nanoscience, Delft University of Technology. During part of this research, M. E. was supported by Postdoctoral Fellowship of Japan Society for Promotion of Science (JSPS) for overseas researchers and JSPS KAKENHI (Grant No. 16F16325).
\end{acknowledgments}

\appendix
%%%%%%%%%%

\section{\label{app1}Hamiltonian}

This section reviews well-known results and defines our notation
\cite{Suhl1957,Bryant1988,Stancil2009,Rezende2009}. The external magnetic
field and equilbrium magnetization are along $\hat{z}$.

\textit{Dipolar interaction}: The (time-dependent) dipolar (Zeeman)
interaction reads
\begin{equation}
H^{(d)}=-\frac{\mu_0}{2}\int\vec{m}\left(  \vec{r}\right)  \cdot\vec{h}%
^{(d)}\left(  \vec{r}\right)  d\vec{r}, \label{Aeq1}%
\end{equation}
where $\mu_0$ is the vacuum permeability, $\vec{h}^{(d)}$ is the dipolar field and $\vec{m}$ is the magnetization
texture in real space. $H^{(d)}$ for a bulk magnet can be derived from a
Heisenberg Hamiltonian for a lattice of $N$ cells with spin $S$ by the
Holstein-Primakoff (HP) transformation. Internal excitations of large spins in
materials such as YIG\ with local moment $S=5/2$ start to play a role only
when the local spin excitation exceeds $\hbar\ $or $n_{0}/(NS)\gtrsim0.1$. We
operate here at $\sim1\,$K, so $n_{0}/(NS)\lesssim0.01$, which implies that
the simple Holstein-Primakoff expansion is valid. The magnetization vector
$\vec{\mathfrak{m}}_{\vec{k}}$ in Fourier space can then be written in terms
of the magnon annihilation (creation) field operators $a_{\vec{k}}$ ($a_{\vec{k}}^{\dag}$). To third order in field operators
\begin{align}
\mathfrak{m}_{x,\vec{k}}  &  =\hbar\gamma\left(  \frac{NS}{2V}\right)
^{\frac{1}{2}}\left(  a_{\vec{k}}+a_{-\vec{k}}^{\dag}\right)  -\hbar
\gamma\left(  \frac{\hbar\gamma}{32M_{s}V^{2}}\right)  ^{\frac{1}{2}%
}\nonumber\\
&  \left(  \sum_{\vec{k}_{1},\vec{k}_{2}}a_{\vec{k}_{1}}^{\dag}a_{\vec{k}_{2}%
}^{\dag}a_{\vec{k}+\vec{k}_{1}+\vec{k}_{2}}+a_{\vec{k}_{1}}a_{\vec{k}_{2}%
}a_{-\vec{k}+\vec{k}_{1}+\vec{k}_{2}}^{\dag}\right)  ,\nonumber\\
\mathfrak{m}_{y,\vec{k}}  &  =-i\hbar\gamma\left(  \frac{NS}{2V}\right)
^{\frac{1}{2}}\left(  a_{\vec{k}}-a_{-\vec{k}}^{\dag}\right)  +i\hbar
\gamma\left(  \frac{\hbar\gamma}{32M_{s}V^{2}}\right)  ^{\frac{1}{2}%
}\nonumber\\
&  \left(  \sum_{\vec{k}_{1},\vec{k}_{2}}a_{\vec{k}_{1}}^{\dag}a_{\vec{k}_{2}%
}^{\dag}a_{\vec{k}+\vec{k}_{1}+\vec{k}_{2}}-a_{\vec{k}_{1}}a_{\vec{k}_{2}%
}a_{-\vec{k}+\vec{k}_{1}+\vec{k}_{2}}^{\dag}\right), \nonumber\\
\mathfrak{m}_{z,\vec{k}}  &  =M_{s}\sqrt{V}-\frac{\gamma\hbar}{\sqrt{V}}\sum_{\vec{k}_1}{a_{-\vec{k}_1}^{\dag}a_{\vec{k}-\vec{k}_1}}, \label{Aeq2}
\end{align}
where $M_{s}=g\mu_{B}NS/V$ is the saturation magnetization, $V$ is the volume
of the sample, $g$ is the electron g-factor, and $\mu_{B}$ is the Bohr magneton.

When $k^{-1}$, with $k=|\vec{k}|,$ approaches the sample dimensions, the
spectrum becomes a discrete manifold of magnetostatic modes. For the uniform
mode
\begin{equation}
\vec{h}^{(d),\left(  0\right)  }=\left(  -N_{x}\mathfrak{m}_{x,\vec{k}=0}%
\hat{x}-N_{y}\mathfrak{m}_{y,\vec{k}=0}\hat{y}-N_{z}\mathfrak{m}_{z,\vec{k}%
=0}\hat{z}\right)  , \label{Aeq3}%
\end{equation}
where for a sphere the demagnetizing constants $N_{x(y,z)}=1/3$, while {for a
thin film, $N_{x}=N_{y}=0$ and $N_{z}=1$. }For large enough $k$, the magnons
in a sphere are well described by plane waves and a continuous spectrum. Their
dipolar field reads
\begin{equation}
\vec{h}^{(d)}=-\sum_{\vec{k}\neq0}\hat{k}\left(  \hat{k}\cdot\mathfrak{\vec
{m}}_{\vec{k}}\right)  , \label{eq4}%
\end{equation}
Following Suhl \cite{Suhl1957}, we use Eq. (\ref{eq4}) for all $\vec{k}\neq0$
when computing magnon interactions. The dipolar interaction Hamiltonian Eqs.
(\ref{Aeq1}), (\ref{Aeq3}) and (\ref{eq4}) can then be written as a sum of
several terms involving the Kittel mode $\delta_{\vec{k},0}$ and plane spin
wave (PW) modes $1-\delta_{\vec{k},0}$,
\begin{align}
 & H^{(d)}   =\sum_{\vec{k}}\frac{1}{2}\left\{  \left[  \frac{|k_{+}|^{2}%
}{2k^{2}}\mathfrak{g}^{2}\left(  1-\delta_{\vec{k},0}\right)  \right]
a_{\vec{k}}^{\dag}a_{\vec{k}}\right. \nonumber\\
&  +\left(  N_{T}^{\prime}\mathfrak{g}^{2}\delta_{\vec{k},0}-2N_{z}%
\mathfrak{g^{\prime\prime}}\mathfrak{g^{\prime\prime\prime}}\right)
a_{\vec{k}}^{\dag}a_{\vec{k}}\nonumber\\
&  +\left(  \left[  \frac{k_{+}^{2}}{4k^{2}}\mathfrak{g}^{2}\left(
1-\delta_{\vec{k},0}\right)  +\frac{N_{T}}{2}\mathfrak{g}^{2}\delta_{\vec
{k},0}\right]  a_{\vec{k}}^{\dag}a_{-\vec{k}}^{\dag}+\mathrm{H.c.}\right)
\nonumber\\
&  +\sum_{\vec{k}^{\prime}}\left(  \left[  \frac{-k_{+}k_{z}}{2k^{2}%
}\mathfrak{g}\mathfrak{g^{\prime\prime\prime}}\left(  1-\delta_{\vec{k}%
,0}\right)  \right]  a_{\vec{k}}^{\dag}a_{\vec{k}^{\prime}}^{\dag}a_{\vec
{k}+\vec{k}^{\prime}}+\mathrm{H.c.}\right) \nonumber\\
&  +\sum_{\vec{k}^{\prime},\vec{k}^{\prime\prime}}\left(  \left[  \frac
{k_{z}^{2}}{k^{2}}\mathfrak{g^{\prime\prime\prime}}^{2}\left(  1-\delta
_{\vec{k},0}\right)  \right]  a_{\vec{k}^{\prime}}^{\dag}a_{\vec{k}%
^{\prime\prime}}^{\dag}a_{\vec{k}+\vec{k}^{\prime}}a_{-\vec{k}+\vec{k}%
^{\prime\prime}}\right) \nonumber\\
&  -\sum_{\vec{k}^{\prime},\vec{k}^{\prime\prime}}\left(  N_{T}\mathfrak{g}%
\mathfrak{g^{\prime}}\delta_{\vec{k},0}a_{0}^{\dag}a_{\vec{k}^{\prime}+\vec
{k}^{\prime\prime}}^{\dag}a_{\vec{k}^{\prime}}a_{\vec{k}^{\prime\prime}%
}+\mathrm{H.c.}\right) \nonumber\\
&  -\sum_{\vec{k}^{\prime},\vec{k}^{\prime\prime}}\left\{  \left[  \frac
{k_{-}^{2}}{k^{2}}\mathfrak{g}\mathfrak{g^{\prime}}\left(  1-\delta
_{\vec{k},0}\right)  \right]  a_{\vec{k}^{\prime}}^{\dag}a_{\vec{k}%
^{\prime\prime}}^{\dag}a_{\vec{k}}a_{-\vec{k}+\vec{k}^{\prime}+\vec{k}%
^{\prime\prime}}+\mathrm{H.c.}\right\} \nonumber\\
&  +\left.  N_{z}\mathfrak{g^{\prime\prime\prime}}^{2}\delta_{\vec{k},0}%
\sum_{\vec{k}^{\prime},\vec{k}^{\prime\prime}}a_{\vec{k}^{\prime}}^{\dag
}a_{\vec{k}^{\prime\prime}}^{\dag}a_{\vec{k}^{\prime}}a_{\vec{k}^{\prime
\prime}}\right\}  , \label{eq5}%
\end{align}
where $\mathfrak{g}=\sqrt{2\hbar\mu_0\gamma M_{s}}$, $\mathfrak{g^{\prime}}%
=\hbar\gamma\sqrt{\hbar\mu_0\gamma/\left(  32M_{s}V^{2}\right)  }$,
$\mathfrak{g^{\prime\prime}}=M_{s}\sqrt{\mu_0V}$, $\mathfrak{g^{\prime\prime\prime
}}=\hbar\gamma\sqrt{\mu_0/V}$, $k_{\pm}=k_{x}\pm ik_{y}$, $2N_{T}=N_{x}-N_{y}$,
$2N_{T}^{\prime}=N_{x}+N_{y}$. In this notation, in order to conserve the units $\gamma$ values should be input in units of $1/(\text{T}\cdot\text{s})$.

\textit{Exchange interaction}: The exchange Hamiltonian in real space
$H^{(exc)}=\mu_0\mathcal{E}/\left(  2M_{s}V\right)  \int\left(  \nabla\vec{m}%
(\vec{r})\right)  ^{2}d\vec{r}$, where $\mathcal{E}$ is the exchange constant.
In momentum space $H^{(exc)}=\mu_0\mathcal{E}/\left(  2M_{s}\right)\sum_{\vec{k},\vec{k}^{\prime}}k^2\vec{\mathfrak{m}}_{\vec{k}}\cdot\vec{\mathfrak{m}}_{\vec{k}^{\prime}}$, which leads to 
\begin{align}
& H^{(exc)}  =\nonumber\\
&  \sum_{\vec{k}\neq0}k^{2}\frac{\mathcal{E}}{2M_{s}}\left\{  \mathfrak{g}%
^{2}a_{\vec{k}}^{\dag}a_{\vec{k}}+\mathfrak{g^{\prime\prime\prime}}^{2}%
\sum_{\vec{k}^{\prime},\vec{k}^{\prime\prime}}\left(  a_{\vec{k}^{\prime}%
}^{\dag}a_{\vec{k}^{\prime\prime}}^{\dag}a_{\vec{k}+\vec{k}^{\prime}}%
a_{-\vec{k}+\vec{k}^{\prime\prime}}\right)  \right\}  .\label{eq6}%
\end{align}

\textit{Crystalline anisotropy}: The crystalline magnetic anisotropy energy
$H^{(A)}=-\mu_0\vec{m}\cdot\bar{N}_{A}\vec{m}$ in terms of the anisotropy tensor
$\bar{N}_{A}.$ We assume here easy-axis or easy-plane anisotropy with crystal
symmetry axis along $\vec{M}_{0}.$ $\bar{N}_{A}$ is then diagonal with
elements $N_{A,x(y,z)}$ and can be classified in terms of symmetric
($2N_{A,s}=N_{A,x}+N_{A,y}$), antisymmetric ($2N_{A,as}=N_{A,x}-N_{A,y}$), and
($N_{A,z}$) components, leading to
\begin{align}
& H^{(A)}   =-\sum_{\vec{k}}\frac{1}{2}\left\{  \left(  \mathfrak{g}^{2}%
N_{A,s}-2\mathfrak{g^{\prime\prime}}\mathfrak{g^{\prime\prime\prime}}%
N_{A,z}\right)  a_{\vec{k}}^{\dag}a_{\vec{k}}+\right.  \nonumber\\
&  \frac{\mathfrak{g}^{2}}{2}N_{A,as}\left(  a_{\vec{k}}^{\dag}a_{-\vec{k}%
}^{\dag}+a_{\vec{k}}a_{-\vec{k}}\right)  \nonumber\\
&  -\sum_{\vec{k}^{\prime},\vec{k}^{\prime\prime}}N_{A,as}\left(
2\mathfrak{g}\mathfrak{g^{\prime}}a_{\vec{k}^{\prime}}^{\dag}a_{\vec
{k}^{\prime\prime}}^{\dag}a_{\vec{k}+\vec{k}^{\prime}+\vec{k}^{\prime\prime}%
}a_{-\vec{k}}+H.c.\right)  +\nonumber\\
&  \left.  \sum_{\vec{k}^{\prime},\vec{k}^{\prime\prime}}N_{A,z}\left(
\mathfrak{g^{\prime\prime\prime}}^{2}a_{\vec{k}^{\prime}}^{\dag}a_{\vec
{k}^{\prime\prime}}^{\dag}a_{\vec{k}+\vec{k}^{\prime}}a_{-\vec{k}+\vec
{k}^{\prime\prime}}\right)  \right\}  .\label{eq7}%
\end{align}
\textit{ }The first term $\frac{1}{2}\left(  \mathfrak{g}^{2}N_{A,s}%
-2\mathfrak{g^{\prime\prime}}\mathfrak{g^{\prime\prime\prime}}N_{A,z}\right)
a_{\vec{k}}^{\dag}a_{\vec{k}}$ causes only a small constant shift of the
dispersion that may be disregarded. In a cubic crystal, when $\vec{M}_{0}%
\Vert\lbrack001]$, $N_{A,s}\neq0$ and $N_{A,as}=N_{z}=0$
\cite{Stancil2009,Hansen1974}, while for $\vec{M}_{0}\Vert\lbrack111]$,
$N_{A,s}=N_{A,as}=0$ and $N_{A,z}\neq0$. When $\vec{M}_{0}\Vert\lbrack110]$,
$N_{A,as}\approx3N_{A,s}$, and $N_{A,z}\approx2N_{A,as}$, the crystal
anisotropy affects the Kittel mode besides a constant shift by $\mathfrak{g}%
^{2}N_{A,as}(a_{0}^{\dag}a_{0}^{\dag}+a_{0}a_{0})/2$, again to lowest
(bilinear) order in the field operators.

\textit{Zeeman interaction}: The Zeeman energy of an applied magnetic field
$\vec{H}_{ext}=H_{ext}\hat{z}\Vert\vec{M}_{0}$
\begin{equation}
H^{(Z)}=\hbar\gamma \mu_0 H_{ext}\sum_{\vec{k}}a_{\vec{k}}^{\dag}a_{\vec{k}}
\label{eq8}%
\end{equation}
depends only on the total number of magnons.

\textit{Total magnetic Hamiltonian}: Collecting Eqs. (\ref{eq5}-\ref{eq8}),
the total Hamiltonian becomes $H^{(T,m)}=\sum_{\vec{k}}H_{\vec{k}}^{(T,m)}$
with
\begin{align}
H_{\vec{k}}^{(T,m)}  &  =\mathcal{A}_{\vec{k}}a_{\vec{k}}^{\dag}a_{\vec{k}%
}+\left[  \mathcal{B}_{\vec{k}}a_{-\vec{k}}^{\dag}a_{\vec{k}}^{\dag
}+\mathrm{H.c}.\right]  +\nonumber\\
&  \sum_{\vec{k}^{\prime}}\left[  \mathcal{C}_{\vec{k}}a_{\vec{k}}^{\dag
}a_{\vec{k}^{\prime}}^{\dag}a_{\vec{k}+\vec{k}^{\prime}}+\mathrm{H.c.}\right]
+\nonumber\\
&  \sum_{\vec{k}^{\prime},\vec{k}^{\prime\prime}}\left[  \mathcal{D}_{\vec{k}%
}a_{\vec{k}^{\prime}}^{\dag}a_{\vec{k}^{\prime\prime}}^{\dag}a_{\vec{k}%
}a_{-\vec{k}+\vec{k}^{\prime}+\vec{k}^{\prime\prime}}+H.c.\right]
+\nonumber\\
&  \sum_{\vec{k}^{\prime},\vec{k}^{\prime\prime}}\left[  \mathcal{D}_{\vec{k}%
}^{\prime}a_{\vec{k}^{\prime}}^{\dag}a_{\vec{k}^{\prime\prime}}^{\dag}%
a_{\vec{k}+\vec{k}^{\prime}}a_{-\vec{k}+\vec{k}^{\prime\prime}}\right]  ,
\label{eq9}%
\end{align}
and matrix elements
\begin{align}
\mathcal{A}_{\vec{k}}  &  =\frac{|k_{+}|^{2}g^{2}}{4k^{2}}\left(
1-\delta_{k,0}\right)  +\frac{1}{2}N_{T}^{\prime}\mathfrak{g}^{2}\delta
_{k,0}-N_{z}g^{\prime\prime}\mathfrak{g}^{\prime\prime\prime}\nonumber\\
&  +\frac{k^{2}\mathcal{E}\mathfrak{g}^{2}}{2M_{s}}-\frac{\mathfrak{g}^{2}}%
{2}N_{A,s}+\mathfrak{g}^{\prime\prime}\mathfrak{g}^{\prime\prime\prime}%
N_{A,z}+\hbar\gamma H_{ext},\label{eq_a10}\\
\mathcal{B}_{\vec{k}}  &  =\frac{k_{+}^{2}g^{2}}{8k^{2}}\left(  1-\delta
_{k,0}\right)  +\frac{N_{T}\mathfrak{g}^{2}}{4}\delta_{k,0}-\frac
{\mathfrak{g}^{2}}{4}N_{A,as},\label{eq_a11}\\
\mathcal{C}_{\vec{k}}  &  =-\frac{k_{+}k_{z}gg^{\prime\prime\prime}}{4k^{2}%
}\left(  1-\delta_{k,0}\right)  ,\label{eq_a12}\\
\mathcal{D}_{\vec{k}}  &  =-\frac{k_{-}^{2}gg^{\prime}}{2k^{2}}\left(
1-\delta_{k,0}\right)  -\frac{1}{2}N_{T}gg^{\prime}\delta_{k,0}+gg^{\prime
}N_{A,as},\label{eq_a13}\\
\mathcal{D}_{\vec{k}}^{\prime}  &  =\frac{k_{z}^{2}\mathfrak{g}^{\prime
\prime\prime2}}{2k^{2}}\left(  1-\delta_{k,0}\right)  +\frac{1}{2}%
N_{z}g^{\prime\prime\prime2}\delta_{k,0}\nonumber\\
&  -\frac{1}{2}N_{A,z}\mathfrak{g}^{\prime\prime\prime2}+\frac{k^{2}%
\mathcal{E}\mathfrak{g}^{\prime\prime\prime2}}{2M_{s}},\label{eq_a14}
\end{align}
The term $\mathcal{B}_{\vec{k}}a_{-\vec{k}}^{\dag}a_{\vec{k}}^{\dag
}+\mathrm{H.c.}$ in Eq. (\ref{eq9}) is diagonalized by the Bogoliubov
transformation
\begin{align}
a_{\vec{k}}  &  =u_{\vec{k}}c_{\vec{k}}-v_{\vec{k}}c_{-\vec{k}}^{\dag
}\nonumber\\
a_{\vec{k}}^{\dag}  &  =u_{\vec{k}}^{\ast}c_{\vec{k}}^{\dag}-v_{\vec{k}}%
^{\ast}c_{-\vec{k}}, \label{eq10}%
\end{align}
with
\begin{equation}
u_{\vec{k}}=\left(  \frac{\mathcal{A}_{\vec{k}}+\omega_{\vec{k}}}%
{2\omega_{\vec{k}}}\right)  ^{\frac{1}{2}};v_{\vec{k}}=\frac{\mathcal{B}%
_{\vec{k}}}{|\mathcal{B}_{\vec{k}}|}\left(  \frac{\mathcal{A}_{\vec{k}}%
-\omega_{\vec{k}}}{2\omega_{\vec{k}}}\right)  ^{\frac{1}{2}}, \label{eq11}%
\end{equation}
and $\omega_{\vec{k}}=\sqrt{\mathcal{A}_{\vec{k}}^{2}-|\mathcal{B}_{\vec{k}%
}|^{2}}$. The quadratic terms in $H^{(T,m)}$ in Eq. (\ref{eq9}) reduce to the
harmonic oscillator $\omega_{\vec{k}}c_{\vec{k}}^{\dag}c_{\vec{k}}$.

After some algebra, the three magnon terms in the second line of Eq.
(\ref{eq9}) may be transformed and simplified to
\begin{equation}
H^{(3MS)}=\sum_{\vec{k}}\left(  \mathcal{C}_{\vec{k}}^{(3MS)}c_{0}c_{\vec{k}%
}^{\dag}c_{-\vec{k}}^{\dag}+H.c.\right)  ,\nonumber
\end{equation}
where
\begin{align}
\mathcal{C}_{\vec{k}}^{(3MS)}  &  =\left[  \mathcal{C}_{\vec{k}}\left(
u_{\vec{k}}^{\ast}v_{0}^{\ast}v_{\vec{k}}+|u_{\vec{k}}|^{2}u_{0}\right)
\right. \nonumber\\
&  \left.  +\mathcal{C}_{\vec{k}}^{\ast}\left(  -v_{\vec{k}}u_{0}u_{\vec{k}%
}^{\ast}-|v_{\vec{k}}|^{2}v_{0}^{\ast}\right)  \right]  .
\end{align}

The four-magnon scattering terms are transformed into $H^{(4MS)}=\sum_{\vec
{k},\vec{k}^{\prime},\vec{k}^{\prime\prime}}\left(  H_{\vec{k},\vec{k}%
^{\prime},\vec{k}^{\prime\prime}}^{(4MS,1)}+H_{\vec{k},\vec{k}^{\prime}%
,\vec{k}^{\prime\prime}}^{(4MS,2)}\right)  $, where $H_{\vec{k},\vec
{k}^{\prime},\vec{k}^{\prime\prime}}^{(4MS,1)}$ corresponds to the third line
of Eq. (\ref{eq9}), while $H_{\vec{k},\vec{k}^{\prime},\vec{k}^{\prime\prime}%
}^{(4MS,2)}$ corresponds to the fourth line of Eq. (\ref{eq9}). Keeping only
the combinations that can satisfy resonant conditions leads to the simplified
\begin{align}
&  H^{(4MS)}  =\nonumber\\
&  \sum_{\vec{k},\vec{k^{\prime}}}\left[  \left(  \mathcal{D}_{\vec{k},\vec
{k}^{\prime}}^{4MS,1}c_{\vec{k}}^{\dag}c_{\vec{k}}c_{\vec{k}^{\prime}}^{\dag
}c_{\vec{k}^{\prime}}+\mathcal{D}_{\vec{k},\vec{k}^{\prime}}^{4MS,2}c_{\vec
{k}}^{\dag}c_{-\vec{k}}^{\dag}c_{\vec{k}^{\prime}}c_{-\vec{k}^{\prime}%
}\right)  +\mathrm{H.c.}\right],\label{eq_a18}
\end{align}
where
\begin{align}
&  \mathcal{D}_{\vec{k},\vec{k}^{\prime}}^{4MS,1}  =\nonumber\\
&  \left\{  2\left[  \mathcal{D}_{\vec{k}}\left(  |u_{\vec{k}}|^{2}|u_{\vec
{k}^{\prime}}|^{2}\right)  +\mathcal{D}_{\vec{k}}^{\ast}\left(  |v_{\vec{k}%
}|^{2}|v_{\vec{k}^{\prime}}|^{2}\right)  \right]  +\right. \nonumber\\
&  2\left[  \mathcal{D}_{\vec{k}}\left(  |u_{\vec{k}}|^{2}|v_{\vec{k}^{\prime
}}|^{2}\right)  +\mathcal{D}_{\vec{k}}^{\ast}\left(  |v_{\vec{k}}|^{2}%
|u_{\vec{k}^{\prime}}|^{2}\right)  \right]  +\nonumber\\
&  \left.  2\left[  \mathcal{D}_{\vec{k}}\left(  u_{\vec{k}^{\prime}}^{\ast
}v_{\vec{k}^{\prime}}^{\ast}u_{\vec{k}}v_{\vec{k}}\right)  +\mathcal{D}%
_{\vec{k}}^{\ast}\left(  u_{\vec{k}^{\prime}}v_{\vec{k}^{\prime}}u_{\vec{k}%
}^{\ast}v_{\vec{k}}^{\ast}\right)  \right]  \right\}  +\nonumber\\
&  \left\{  \left[  \mathcal{D}_{\vec{k}}\left(  |u_{\vec{k}}|^{4}\right)
+\mathcal{D}_{\vec{k}}^{\ast}\left(  |v_{\vec{k}}|^{4}\right)  \right]
+\right. \nonumber\\
&  \left.  2\left[  \mathcal{D}_{\vec{k}}\left(  |u_{\vec{k}}|^{2}|v_{\vec{k}%
}|^{2}\right)  +\mathcal{D}_{\vec{k}}^{\ast}\left(  |u_{\vec{k}}|^{2}%
|v_{\vec{k}}|^{2}\right)  \right]  \right\}  (\delta_{\vec{k}^{\prime},\vec
{k}}+2\delta_{\vec{k}^{\prime},-\vec{k}})+\nonumber\\
&  \mathcal{D}_{\vec{k}}^{\prime}\left[  4u_{0}^{\ast}v_{0}^{\ast}u_{\vec{k}%
}v_{\vec{k}}+|u_{0}|^{2}|u_{\vec{k}}|^{2}+|v_{0}|^{2}|v_{\vec{k}}|^{2}+\right.
\nonumber\\
&  \left.  |u_{0}|^{2}|v_{\vec{k}}|^{2}+|v_{0}|^{2}|u_{\vec{k}}|^{2}\right]
\delta_{\vec{k}^{\prime},0}+\nonumber\\
&  \mathcal{D}_{\vec{k}}^{\prime}\left[  |u_{0}|^{2}|u_{\vec{k}^{\prime}}%
|^{2}+|v_{0}|^{2}|v_{\vec{k}^{\prime}}|^{2}+|u_{0}|^{2}|v_{\vec{k}^{\prime}%
}|^{2}+|v_{0}|^{2}|u_{\vec{k}^{\prime}}|^{2}\right]  {\delta_{\vec{k},0}%
},\label{eq_a19}\\
&  \mathcal{D}_{\vec{k},\vec{k}^{\prime}}^{4MS,2}=\left\{  2\left[
\mathcal{D}_{\vec{k}}\left(  u_{\vec{k}^{\prime}}^{\ast2}u_{\vec{k}}%
^{2}\right)  +\mathcal{D}_{\vec{k}}^{\ast}\left(  v_{\vec{k}^{\prime}}^{\ast
2}v_{\vec{k}}^{2}\right)  \right]  +\right. \nonumber\\
&  \left.  2\left[  \mathcal{D}_{\vec{k}}\left(  u_{\vec{k}^{\prime}}^{\ast
}v_{\vec{k}}^{\ast}u_{\vec{k}}v_{\vec{k}^{\prime}}\right)  +\mathcal{D}%
_{\vec{k}}^{\ast}\left(  v_{\vec{k}^{\prime}}u_{\vec{k}}v_{\vec{k}}^{\ast
}u_{\vec{k}^{\prime}}^{\ast}\right)  \right]  \right\}  +\nonumber\\
&  \mathcal{D}_{\vec{k}}^{\prime}\left[  u_{0}^{\ast2}u_{\vec{k}}^{2}%
+v_{0}^{\ast2}v_{\vec{k}}^{2}+2u_{0}^{\ast}v_{0}u_{\vec{k}}v_{\vec{k}}^{\ast
}\right]  \delta_{\vec{k}^{\prime},0}+\nonumber\\
&  \mathcal{D}_{\vec{k}}^{\prime}\left[  2u_{0}^{\ast}v_{0}u_{\vec{k}^{\prime
}}v_{\vec{k}^{\prime}}^{\ast}\right]  \delta_{\vec{k},0}. \label{eq_a20}%
\end{align}

The system Hamiltonian can be summarized as
\begin{equation}
H^{(T)}=H^{(c)}+H^{(d)}+H^{(mc)}+H^{(T,m)},
\end{equation}
where $H^{(c)}=\omega_{c}b^{\dag}b$ is the cavity photon Hamiltonian,
$H^{(d)}=i\bar{B}(b^{\dag}-b)+\mathrm{H.c.}$ is the external drive of the
cavity field, $\bar{B}=\sqrt{\zeta_{c,ex}P_{in}/(\hbar\omega_{d})}$, coupled
to the cavity by $\zeta_{c,ex}$ via the input mirror, $P_{in}$ is the input
power, $\omega_{d}$ is the input drive frequency, $H^{(mc)}=-iD_{0}(b^{\dag
}c_{0}-bc_{0}^{\dag})$, $D_{0}$ being the coupling constant of the cavity photon
mode and the Kittel mode, and
\begin{equation}
H^{(T,m)}=H^{(3MS)}+H^{(4MS)}+\sum_{\vec{k}}\omega_{\vec{k}}c_{\vec{k}}^{\dag
}c_{\vec{k}}.
\end{equation}

%%%%%%%%%%

\section{\label{app2}Fluctuations}

We can write the field operators as $c_{0}=\alpha_{0}+\delta c_{0}$,
$c_{\vec{k}}=\alpha_{\vec{k}}+\delta c_{\vec{k}}$ and $b=\beta+\delta b$,
where $\{\delta c_{0},\delta c_{\vec{k}},\delta c_{0}\}$ are the fluctuations
around the steady state characterized by the complex numbers $\{\alpha
_{0},\alpha_{\vec{k}},\beta\}$. We define the operators $\delta x_{0(\pm
\vec{k})}=\left[  \delta c_{0(\pm\vec{k})}^{\dag}+\delta c_{0(\pm\vec{k}%
)}\right]  /2$, $\delta y_{0(\pm\vec{k})}=i\left[  \delta c_{0(\pm\vec{k}%
)}^{\dag}-\delta c_{0(\pm\vec{k})}\right]  /2$, $\delta X=\left[  \delta
b^{\dag}+\delta b\right]  /2$, $\delta Y=i\left[  \delta b^{\dag}-\delta
b\right]  /2$, $\mathbf{v}=\left({\delta x_{0}},{\delta p_{0}},{\delta x}_{\vec{k}},{\delta p}_{\vec{k}},{\delta x}_{-\vec{k}},{\delta p}_{-\vec{k}},{\delta X},{\delta Y}\right)^T$, which obey the linearized equation of motion (EOM)
\begin{equation}
\dot{\mathbf{v}}=\mathcal{O}\mathbf{v}+\mathbf{c}. \label{eq30}%
\end{equation}
$\mathbf{c}$ is the vector of the stochastic sources (discussed in more detail
below)
\begin{align}
\mathbf{c}  &  =\left[  \sqrt{\zeta_{mm,0}}F_{mm,0}^{(x)}(t)+\sqrt
{\zeta_{mp,0}}F_{mp,0}^{(x)}(t),\right. \nonumber\\
&  \sqrt{\zeta_{mm,0}}F_{mm,0}^{(p)}(t)+\sqrt{\zeta_{mp,0}}F_{mp,0}%
^{(p)}(t),\nonumber\\
&  \sqrt{\zeta_{mm,\vec{k}}}F_{mm,\vec{k}}^{(x)}(t)+\sqrt{\zeta_{mp,\vec{k}}%
}F_{mp,\vec{k}}^{(x)}(t),\nonumber\\
&  \sqrt{\zeta_{mm,\vec{k}}}F_{mm,\vec{k}}^{(p)}(t)+\sqrt{\zeta_{mp,\vec{k}}%
}F_{mp,\vec{k}}^{(p)}(t),\nonumber\\
&  \sqrt{\zeta_{mm,-\vec{k}}}F_{mm,-\vec{k}}^{(x)}(t)+\sqrt{\zeta_{mp,-\vec
{k}}}F_{mp,-\vec{k}}^{(x)}(t),\nonumber\\
&  \sqrt{\zeta_{mm,-\vec{k}}}F_{mm,-\vec{k}}^{(p)}(t)+\sqrt{\zeta_{mp,-\vec
{k}}}F_{mp,-\vec{k}}^{(p)}(t),\nonumber\\
&  \sqrt{\zeta_{c,0}}F_{c,0}^{(x)}(t)+\sqrt{\zeta_{c,ex}}F_{c,ex}%
^{(x)}(t),\nonumber\\
&  \left.  \sqrt{\zeta_{c,0}}F_{c,0}^{(p)}(t)+\sqrt{\zeta_{c,ex}}%
F_{c,ex}^{(p)}(t)\right]  ,
\end{align}
where $F_{mm(mp),0(\pm\vec{k})}^{(x)}(t)=[F_{mm(mp),0(\pm\vec{k})}^{\dag
}(t)+F_{mm(mp),0(\pm\vec{k})}(t)]/2$, $F_{mm(mp),0(\pm\vec{k})}^{(p)}%
(t)=i[F_{mm(mp),0(\pm\vec{k})}^{\dag}(t)-F_{mm(mp),0(\pm\vec{k})}(t)]/2$,
$F_{c,0(ex)}^{(x)}(t)=[F_{c,0(ex)}^{\dag}(t)+F_{c,0(ex)}(t)]/2$, and
$F_{c,0(ex)}^{(p)}(t)=i[F_{c,0(ex)}^{\dag}(t)-F_{c,0(ex)}(t)]/2$.
$\mathcal{O}$ is a square matrix that is governed by Heisenberg's equation for
the Hamiltonian derived above and serves as well for the stability analysis.
The symmetrized covariance matrix $\boldsymbol{\Lambda}$ consists of equal
time correlations $\left\langle v_{i}v_{j}+v_{j}v_{i}\right\rangle /2$, where
$v_{i(j)}$ is $i$'th ($j$'th) element of $\mathbf{v}$, and
contains the essential statistical parameters. Its EOM $\boldsymbol{\dot
{\Lambda}}=\mathcal{O}\boldsymbol{\Lambda}+\boldsymbol{\Lambda}\mathcal{O}%
^{T}+\boldsymbol{\Gamma,}$ where $\boldsymbol{\Gamma}=\mathbf{c}\mathbf{c}%
^{T}$. Therefore, the steady state covariance matrix $\boldsymbol{\Lambda
}_{\infty}$ is solution of the linear system of equation $\mathcal{O}%
\Lambda_{\infty}+\Lambda_{\infty}\mathcal{O}^{T}+\Gamma=0$.

Rotating $\delta X$ ($\delta Y$) by an angle $\theta$ to a
new variable $\delta X_{\theta}=(\delta b^{\dag}e^{i\theta}+\delta
be^{-i\theta})/2$ [$\delta Y_{\theta}=i(\delta b^{\dag}e^{i\theta}-\delta
be^{-i\theta})/2$] leads to a steady state covariance matrix $\Lambda
_{\infty,\theta}$. We need the ellipticity of the total field fluctuation,
i.e. the elements $[\Lambda_{\infty,\theta}]_{7(8),7(8)}$.

Let $(\alpha_{0},\alpha_{\pm\vec{k}},\beta)$ be the steady-state mean-field
solutions of the EOMs (\ref{eq34_new})-(\ref{eq36_new}). The matrix
$\mathcal{O}$ follows by linearizing these equations without noise terms
around the steady state. For example, a term of the form $c_{0}^{\dag}%
c_{0}c_{0}$ is linearized as $2|\alpha_{0}|^{2}\delta c_{0}+\alpha_{0}%
^{2}\delta c_{0}^{\dag}$, where $\alpha_{0}$ is the mean field solution. The
covariance matrix can then be computed as explained above. The fluctuation
ellipse of the total field is parameterized by
\begin{equation}
\xi_{sq}=\frac{\min\left\{  [\Lambda_{\infty,\theta}]_{7(8),7(8)}\right\}
}{\max\left\{  [\Lambda_{\infty,\theta}]_{7(8),7(8)}\right\}  },\label{eq32}%
\end{equation}
as well as $\theta_{sq}$ which is $\theta$ where $\left\{  [\Lambda
_{\infty,\theta}]_{7,7}\right\}  $ is minimum.
%%%%%%%%%%

\subsection{\label{app2_1}4MS effects}

In the absence of microwaves, the \textquotedblleft
self-Kerr\textquotedblright\ term $K_{1}(c_{0}^{\dag}c_{0})^{2}$ of the Kittel
mode, with $K_{1}=2\operatorname{Re}\left[  \mathcal{D}_{0,0}^{4MS,1}%
+\mathcal{D}_{0,0}^{4MS,2}\right]  $, drives an initially coherent state
through a cycle of periodic collapses and revivals in phase space, during
which squeezed coherent states \cite{Walls2008} and nonclassical
superpositions of two or more coherent states in phase space develop
\cite{Kirchmair2013}. The Kittel coherent state $|\alpha_{0}\rangle$ can be
expanded in number states $|n\rangle$ as $|\alpha_{0}\rangle=\exp
({-|\alpha_{0}|^{2}/2})\sum_{n}\alpha_{0}^{n}/\sqrt{n!}|n\rangle$. Ignoring
dissipation, the temporal evolution of this state reads $|\Psi(t)\rangle
=\exp({-K_{1}(c_{0}^{\dag}c_{0})^{2}t})|\alpha_{0}\rangle=\exp({-|\alpha
_{0}|^{2}/2})\sum_{n}(\alpha_{0}\exp(-K_{1}nt))^{n}/\sqrt{n!}|n\rangle$. $|\Psi(t)\rangle=|\alpha_{0}\rangle$ \textquotedblleft revives\textquotedblright\ at $t=m\pi/K_{1},$ where $m\geq0$ is integer. At
$t=\pi/(mK_{1})$ superposition \textquotedblleft Schr\"{o}dinger
cat\textquotedblright\ states develop, for example $\Psi(\pi/(2K_{1}%
))=1/\sqrt{2}(\exp({-i\pi/4})|\alpha_{0}\rangle+\exp({i\pi/4})|-\alpha
_{0}\rangle)$, where\textit{ }$|-\alpha_{0}\rangle$ is a Kittel coherent state
with opposite phase. These processes cannot develop when the self-Kerr
coefficient is small compared to the damping. However, the steady state in the
presence of a constant coherent microwave drive remains coherent, but is
\textquotedblleft squeezed\textquotedblright\ by the non-linear terms as
explained in the following.

The fluctuation ellipse of the Kittel mode is affected as well by its
instability into a coherent superposition of the $\vec{k}$ and $-\vec{k}$
modes via the interaction $c_{\vec{k}}^{\dag}c_{-\vec{k}}^{\dag}c_{0}c_{0}$,
leading to finite $\langle c_{\vec{k}}^{\dag}c_{-\vec{k}}^{\dag}\rangle$ and
thereby $|\alpha_{\vec{k}}|^{2}\left(  e^{-i(\phi_{\vec{k}}+\phi_{-\vec{k}}%
)}c_{0}c_{0}+e^{i(\phi_{\vec{k}}+\phi_{-\vec{k}})}c_{0}^{\dag}c_{0}^{\dag
}\right)  $, where $\phi_{\pm\vec{k}}$ is the phase of $\alpha_{\pm\vec{k}}$.
With mean-field steady state of the Kittel mode $\alpha_{0}$, the effective
Hamiltonian (after integrating out the cavity field) up to second order in
$\delta c_{0}$ ($\delta c_{0}^{\dag}$), and ignoring fluctuations $\delta
c_{\pm\vec{k}}$ ($\delta c_{\pm\vec{k}}^{\dag}$) in $\vec{k}\neq0$, reads
\begin{align}
H^{eff}  &  =\Delta_{0}^{\prime}\delta c_{0}^{\dag}\delta c_{0}+\left[
(K_{s}+\mathcal{G})\delta c_{0}^{\dag}\delta c_{0}^{\dag}+H.c.\right]
+\nonumber\\
&  \left[  (\Delta_{0}^{\prime}+K_{1}|\alpha_{0}|^{2}\alpha_{0}+\mathcal{G}\alpha_{0}^{\ast}+B^{\prime\ast})\delta c_{0}^{\dag}+H.c.\right]  ,
\label{eq34}%
\end{align}
where $\Delta_{0}^{\prime}=\{\Delta_{0}-(D_{0}^{2}\Delta)/(\Delta^{2}+\zeta
_{c}^{2}/4)+4\operatorname{Re}[\mathcal{D}_{0,\vec{k}}^{4MS,1}]|\alpha
_{\vec{k}}|^{2}\}$, $\mathcal{G}=\mathcal{D}_{0,\vec{k}}^{4MS,2}|\alpha
_{\vec{k}}|^{2}\exp[i(\phi_{\vec{k}}+\phi_{-\vec{k}})]$, $B^{\prime}=(-i\Delta
ED_{0})/(\Delta^{2}+\zeta_{c}^{2}/4)$, and $K_{s}=|\alpha_{0}|^{2}%
e^{2i\phi_{0}}$, where $\phi_{0}$ is the phase of $\alpha_{0}$. Since $\delta
c_{0}^{\dag}$ and $\delta c_{0}$ in the last term vanish when operating on
$\alpha_{0}$, to leading order
\begin{equation}
H^{eff}=\Delta_{0}^{\prime}\delta c_{0}^{\dag}\delta c_{0}+\left[
(K_{s}+\mathcal{G})\delta c_{0}^{\dag}\delta c_{0}^{\dag}+\mathrm{H.c.}%
\right]  . \label{eq44_new2}%
\end{equation}

We may diagonalize Eq. (\ref{eq44_new2}) by the Bogoliubov transformation
$\delta c_{0}^{\prime}=u_{0}\delta c_{0}-v_{0}^{\ast}\delta c_{0}^{\dag}$,
which acts on the vacuum as a squeezing operator $\Pi(\epsilon)=\exp\left[
1/2\epsilon^{\ast}c_{0}^{2}-1/2\epsilon c_{0}^{\dag2}\right]  $, i.e. $\delta
c_{0}^{\prime}=\Pi(\epsilon)\delta c_{0}\Pi^{\dag}(\epsilon)$, where
$\epsilon=-v_{0}\tanh^{-1}(|v_{0}|/u_{0})/|v_{0}|$. $\epsilon=r_{s}%
e^{2i\theta_{s}}$ parameterizes the fluctuation ellipse: $e^{r_{s}}$
($e^{-r_{s}}$) is the major (minor) diameter and $\theta_{s}$ is the angle of
the major axis of the ellipse.

The squeezing parameters $r_{s}$ and $\theta_{s}$ are functions of $K_{s}$ and
$\mathcal{G}$. When the Kittel mode is stable, $\left\vert \alpha_{\vec{k}%
}\right\vert ^{2}=\mathcal{G}=0$.\textit{ }At the instability threshold of the
Kittel mode, a pair of oppositely moving magnons with momenta $\vec{k}$ and
$-\vec{k}$ is excited and $|\alpha_{\vec{k}}|^{2}\neq0$. $\mathcal{G}$ then
may grow to become of the same order of magnitude as $K_{s}$, causing
substantial changes in $r_{s}$ and $\theta_{s}$. It should be noted that the concept of squeezing was first introduced in spintronics in the context of squeezed magnon mediated spin transport \cite{Kamra2016}.

%%%%%%%%
\begin{figure}[!]
\includegraphics[width=0.5\textwidth]{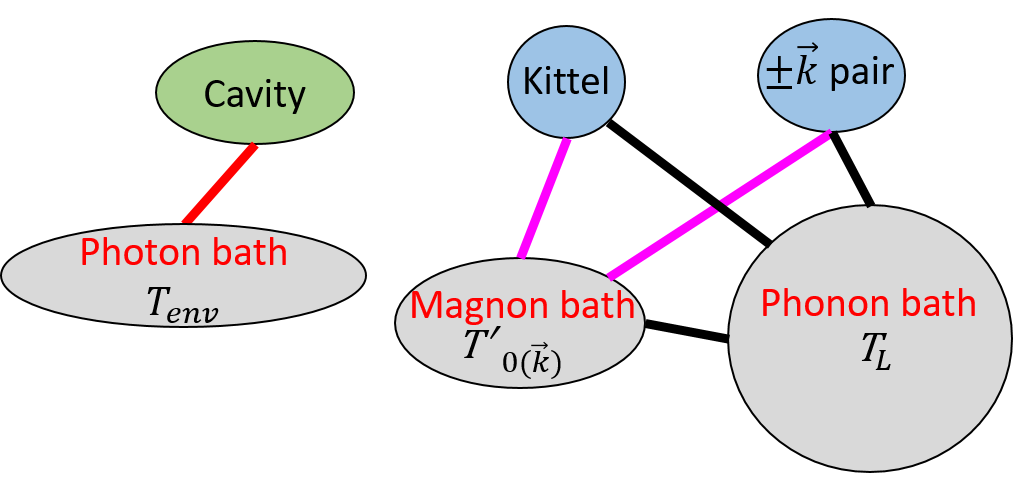}\caption{Schematic of
the total Hilbert space of photonic cavity mode, Kittel mode and selected pair
of magnons with wave vector $\pm\vec{k}$, the baths, and the interactions. The
driven magnetic modes relax by interaction with a bath of thermalized magnons
at temperature $T_{0(\pm\vec{k})}^{\prime}$ as depicted by purple lines as
well as a (phonon) bath corresponding to Geilbert damping at temperature
$T_{L}$ (black lines). The bath of thermalized magnons also is in
contact with the phonon bath. The cavity mode is in contact with a bath at
temperature $T_{\text{env}}$, isolated from baths for magnon modes.}%
\label{fig_SI_2_1}%
\end{figure}

\subsection{\label{app2_2}Baths}

The interaction of the driven state with thermalized magnons
\cite{Rezende2009} and phonons \cite{Ruckriegel2015,Streib2019} govern the parts of
stochastic force matrix $\Gamma$ corresponding to the Kittel mode and the $\pm k\neq 0$ magnon pair. We disregard heating of the phonon bath
which has a much larger specific heat than the magnon system. For a discussion
on heat management of the phonon bath, see also Sec. \ref{sec6}. According to
the fluctuation-dissipation theorem for thermal equilibrium and assuming
temperatures to be high compared to the mode broadening, $\langle F_{mp,0(\pm\vec
{k})}(t),F_{mp,0(\pm\vec{k})}^{\dag}(t^{\prime})\rangle=\zeta_{mp,0(\pm\vec
{k})}(n_{th,0(\pm\vec{k})}+1)\delta(t-t^{\prime})$ and $\langle F_{mp,0(\pm
\vec{k})}^{\dag}(t),F_{mp,0(\pm\vec{k})}(t^{\prime})\rangle=\zeta
_{mp,0(\pm\vec{k})}n_{th,0(\pm\vec{k})}\delta(t-t^{\prime})$, where
$\zeta_{mp,0(\pm\vec{k})}\approx\alpha_{G}\omega_{0(\pm\vec{k})}$ is the
phonon mediated dissipation of magnons, $\alpha_{G}$ the Gilbert damping
constant, {$n_{th,0(\pm\vec{k})}^{-1}=e^{\hbar\omega_{0(\pm\vec{k})}/\left(
k_{B}T_{L}\right)  }-1$ the Planck distribution, }$T_{L}$ the phonon bath
temperature,{ and $k_{B}$ the Boltzmann constant} \cite{Carmichael1999}. The
cavity field is assumed to be in contact with baths that {keeps it at ambient
temperature $T_{\mathrm{env}}$. Therefore,} $\langle F_{c,0(ex)}%
(t)F_{c,0(ex)}^{\dag}(t^{\prime})\rangle=\zeta_{c,0(ex)}(n_{th,\mathrm{env}%
}+1)\delta(t-t^{\prime})$ and $\langle F_{c,0(ex)}^{\dag}(t)F_{c,0(ex)}%
(t^{\prime})\rangle=\zeta_{c,0(ex)}n_{th,\mathrm{env}}\delta(t-t^{\prime})$
[$\zeta_{c,0(ex)}$ is the decay rate of the cavity field by scattering to
internal cavity modes (external leakage), and $n_{th,\mathrm{env}}%
^{-1}=e^{\hbar\omega_{c}/k_{B}T_{\mathrm{env}}}-1$]. The noise sources
$F_{mm,0}(t)$ and $F_{mm,\pm\vec{k}}(t)$ are generated by{\ thermal magnon
bath at temperatures discussed in the following}.

The 4MS driven modes with momenta $0\left(  \pm\vec{k}\right)  $ around
frequency $\omega_{0}$ can relax to other magnon modes by magnon-magnon
scatterings. We can estimate the temperature of the associated thermal cloud
$T_{0(\pm\vec{k})}^{\prime}$ by considering the transition probability
$\mathcal{P}_{p}$ \cite{White1963,White2006}
\begin{equation}
\mathcal{P}_{p}=\frac{2\pi}{\hbar}\int|\langle\Phi^{\prime}|H_{t}|\Phi
\rangle|^{2}\varrho({E})\delta({E^{\prime}}-{E})d{E}, \label{eq35_1}%
\end{equation}
where the initial state $|\Phi\rangle={\prod\limits_{\vec{k}^{\prime}}%
}|n_{\vec{k}^{\prime}}\rangle$ ($\vec{k}^{\prime}=\left\{  {0,}\pm\vec
{k}\right\}  $) and the final state $|\Phi^{\prime}\rangle={\prod
\limits_{\vec{k}^{\prime\prime}}}|n_{\vec{k}^{\prime\prime}}\rangle$ ($\vec
k^{\prime\prime}$ corresponds to thermal magnon bath modes) are expressed in
the magnon number basis with the density of states $\varrho({E})$. The
cross-Kerr term matrix elements in the transition Hamiltonian $H_{t}$ vanish
and elastic three-magnon processes are weak under the condition $H_{ext}\gg
M_{s}/3$ (the assumption in our work){, which leads to $H_{t}=\sum_{\vec
{k}^{\prime\prime}}\mathcal{D}_{\vec{k}^{\prime\prime},\vec{k}^{\prime}%
}^{4MS,2}c_{-\vec{k}^{\prime\prime}}^{\dag}c_{\vec{k}^{\prime\prime}}^{\dag
}c_{-\vec{k}^{\prime}}c_{\vec{k}^{\prime}}+\mathrm{H.c.}$. The scattering rate
of driven magnons into $\vec{k}^{\prime\prime}$ magnon modes
\begin{align}
\mathcal{T}_{sc}  &  =\sum_{\vec{k}^{\prime\prime}}\sum_{\vec{k}^{\prime}%
}2\hbar\omega_{\vec{k}^{\prime}}\left[  \mathcal{P}_{p}\left(  n_{\vec
{k}^{\prime}}\rightarrow n_{\vec{k}^{\prime}}+\delta_{\vec{k}^{\prime},\pm
\vec{k}}+2\delta_{\vec{k}^{\prime},0}\right)  \right. \nonumber\\
&  \left.  -\mathcal{P}_{p}\left(  n_{\vec{k}^{\prime}}\rightarrow n_{\vec
{k}^{\prime}}-\delta_{\vec{k}^{\prime},\pm\vec{k}}-2\delta_{\vec{k}^{\prime
},0}\right)  \right] \nonumber\\
&  =\sum_{\vec{k}^{\prime\prime}}\sum_{\vec{k}^{\prime}}4\pi\omega_{\vec
{k}^{\prime}}\varrho({\hbar\omega_{\vec{k}^{\prime\prime}}})\left\vert
\hbar\mathcal{D}_{\vec{k}^{\prime\prime},\vec{k}^{\prime}}^{4MS,2}\right\vert
^{2}\left[  n_{\vec{k}^{\prime}}^{2}\left(  2n_{\vec{k}^{\prime\prime}%
}+1\right)  +\right. \nonumber\\
&  \left.  n_{\vec{k}^{\prime}}n_{\vec{k}^{\prime\prime}}(6\delta_{\vec
{k}^{\prime},0}+4\delta_{\vec{k}^{\prime},\pm\vec{k}})+n_{\vec{k}%
^{\prime\prime}}(4\delta_{\vec{k}^{\prime},0}+2\delta_{\vec{k}^{\prime}%
,\pm\vec{k}})+\right. \nonumber\\
&  \left.  n_{\vec{k}^{\prime}}(3\delta_{\vec{k}^{\prime},0}+2\delta_{\vec
{k}^{\prime},\pm\vec{k}})+n_{\vec{k}^{\prime\prime}}^{2}(2n_{\vec{k}^{\prime}%
}+1)\times(2\delta_{\vec{k}^{\prime},0}+\delta_{\vec{k}^{\prime},\pm\vec{k}%
})\right]  , \label{eq45_new1}%
\end{align}
where $\mathcal{P}_{p}\left(  n_{\vec{k}^{\prime}}\rightarrow n_{\vec
{k}^{\prime}}\pm\delta_{\vec{k}^{\prime},\pm\vec{k}}\pm2\delta_{\vec
{k}^{\prime},0}\right)  $ corresponds to a transition described by Eq.
(\ref{eq35_1}) that changes magnon number in $\vec{k}^{\prime}$ sector by two.
In the steady state, the scattered magnon flux into the $\vec{k}^{\prime
\prime}$ modes equals their dissipation rate to the lattice
\begin{equation}
\mathcal{T}_{d}=\sum_{\vec{k}^{\prime\prime}}{\hbar\omega_{\vec{k}%
^{\prime\prime}}\zeta_{mp,\vec{k}^{\prime\prime}}}\left(  n_{\vec{k}%
^{\prime\prime}}-n_{th,\vec{k}^{\prime\prime}}\right)  , \label{eq46_new1}%
\end{equation}
{where }$n_{th,\vec{k}^{\prime\prime}}${ is the thermal equilibrium determined
by the phonon bath temperature $T_{L}$}. In order to estimate $n_{\vec
{k}^{\prime\prime}}$ of the magnon bath after heating by the driven magnons,
we assume ${\vec{k}^{\prime\prime}}$ close to ${\vec{k}}$ that dominate
$|\mathcal{D}_{\vec{k}^{\prime\prime},0}^{4MS,2}|$, and assume $\omega
_{\vec{k}^{\prime\prime}}=\omega_{0}=\omega_{\pm\vec{k}^{\prime}}$. These
assumptions also lead to $T_{\vec{k}^{\prime\prime}}=T_{0}$, i.e.
$n_{th,\vec{k}^{\prime\prime}}=n_{th,0}$, $\zeta_{mp,\vec{k}^{\prime\prime}%
}=\zeta_{mp,0}$, and $\varrho({\hbar\omega_{\vec{k}^{\prime\prime}}}%
)=2/(\pi\hbar\zeta_{mp,0})$. For a YIG sphere with $0.1\,\text{mm}$ radius
$|\mathcal{D}_{\vec{k}^{\prime\prime},0}^{4MS,2}|\sim10^{-8}$, and for
$\bar{B}=3.3\times10^{13}$ that we use mainly for the results of this work, the
largest steady states after instabilities $n_{0}\sim n_{\pm\vec{k}}\sim
10^{13}$ [see e.g. Fig. \ref{fig1}(c) and Fig. \ref{fig2}]. Therefore, we can estimate the maximal
heating of the magnon cloud bath in our calculations by equating the
integrands in Eq. (\ref{eq45_new1}) and Eq. (\ref{eq46_new1}). A phonon bath
of $T_{L}=1\,$K, $\omega_{\vec{k}^{\prime\prime}}=\omega_{0}=10^{11}/(2\pi)\,$1/s,
i.e. $n_{th,\vec{k}^{\prime\prime}}=0.87$, $\zeta_{mp,0}= 1\,\text{MHz}$,
determines the mean $\vec{k}^{\prime\prime}$ magnon number to be $n_{\vec
{k}^{\prime\prime}}=1.13$ or a temperature $T_{0({\pm\vec{k}})}^{\prime}%
\sim1.2\,$K, i.e. the magnon modes of the thermal cloud are heated by 0.2
degrees. The correlators of magnon thermal cloud noise source $\langle
F_{mm,0(\pm\vec{k})}(t),F_{mm,0(\pm\vec{k})}^{\dag}(t^{\prime})\rangle
=\zeta_{mm,0(\pm\vec{k})}(n_{th,0(\pm\vec{k})}^{\prime}+1)\delta(t-t^{\prime
})$ and $\langle F_{mm,0(\pm\vec{k})}^{\dag}(t),F_{mm,0(\pm\vec{k})}%
(t^{\prime})\rangle=\zeta_{mm,0(\pm\vec{k})}n_{th,0(\pm\vec{k})}^{\prime
}\delta(t-t^{\prime})$, where $1/n_{0(\pm\vec{k})}^{\prime}=e^{\hbar
\omega_{0(\pm\vec{k})}/k_{B}T_{0(\pm\vec{k})}^{\prime}}-1$, and in YIG
$\zeta_{mm,0(\pm\vec{k})}\sim1$ MHz \cite{Araujo1974,Rezende2009}. }

Schematically, the driven modes and the baths are linked as in Fig.
\ref{fig_SI_2_1}. As temperature goes towards zero, the reservoir correlation time ($\hbar/k_BT$) reaches infinity, and the Markovian bath
approximation breaks down and the noise correlation functions can not be assumed as delta function anymore (i.e. the noise becomes colored) \cite{Carmichael1999}. Morover, the
dominant magnon dissipation source at temperatures $<1\,\text{K}$ is not the
phonon bath anymore \cite{Tabuchi2014}. Therefore, the present approximations
hold for temperatures {$T_{\mathrm{env}}$}$\geq1\,\text{K}$. In our
calculations, we need to set only $T_{\text{env}}$, which determines
$T_{L}\approx T_{\mathrm{env}}$, and consequently $T_{0(\pm\vec{k})}^{\prime}$
which is governed by the rate equations discussed earlier. It should again be
pointed out that realistically, as discussed in Sec. \ref{sec6}, $T_{L}\neq
T_{\mathrm{env}}$, but with proper heat management the difference can be kept
small.
%%%%%%%

\section{\label{app3}Scaling and master equation}

{The observable consequences of non-classical behavior such as entanglement
can be most reliably assessed by the density matrix calculated from the first
principles of quantum mechanics, which is possible by limiting the Hilbert
space of the total Hamiltonian $H^{(T)}$. This can be done by scaling down the
drive amplitude $\bar{B}$ with a coefficient $\mathcal{Q}$ as $\bar
{B}/\mathcal{Q}$},{\ while scaling up the fourth order terms by $\mathcal{Q}%
^{2}$, in order to preserve the non-linearities. }This scaling compresses, but
preserves the details of the phase space, such as the number of fixed points
and their relative positions. The costs of the scaling are loss of transient
states that in the physical system might appear as steady states. For example,
in the scaled system, we never find the limit cycle solution for the Kittel
mode predicted by the (semi-)classical method, see Fig. \ref{fig2}, because the
effects of quantum fluctuations are enhanced by the reduced distance between
the attractors in phase space. Actually, quantum fluctuations always destroy
the classical bistability in the self-Kerr Hamiltonian, but  on very long time
scales when energy minima are well separated \cite{Drummond1980}. In the
off-resonant regime {$\left\vert \omega_{d}-\omega_{c}\right\vert \gg D_{0}$},
we may adiabatically remove (integrate out) the cavity field, which reduces
Hilbert space to the Kittel mode and connected $\pm\vec{k}$ magnon pair
$n_{F,0}\times n_{F,\vec{k}}^{2}$, where $n_{F,0}$ ($n_{F,\vec{k}}$) is the
number of Fock (number) states into which we expand the steady state density
matrix \cite{Carmichael1999,Walls2008}. As an example of how the scaling
preserves the nonlinearity effects, one can note that the bistable points for
the Kittel mode in the absence of 4MS are at $n_{0}^{\pm}=[-2\Delta
_{0}^{\prime}\pm(\Delta_{0}^{\prime2}-3\zeta_{m,0}^{2})^{1/2}]/(6K_{1})$
\cite{Drummond1980}, where $n_{0}^{\pm}$ is the mean number of Kittel mode
magnons in the two bistable points, $\Delta_{0}^{\prime}$ is the effective
detuning of the Kittel mode from the drive frequency, and $K_{1}%
=2\operatorname{Re}[\mathcal{D}_{0,0}^{4MS,1}+\mathcal{D}_{0,0}^{4MS,2}]$.
This number is too large for an exact computation, but we may scale
$n_{0}^{\pm}$ down by a factor $1/\mathcal{Q}^{2}$. We keep dissipation
constant during scaling. This is the scaling that we use in our analysis. One
could alternatively scale the dissipation by $1/\mathcal{Q}$, for which the
detunings at which bistability emerges scale like $1/\mathcal{Q}$, while the
4MS coefficients should be scaled by $\mathcal{Q}$ rather than $\mathcal{Q}%
^{2}$.

We chose $\mathcal{Q}$ which allows us to shrink the Hilbert space to
$n_{F,0}=15$ and $n_{F,\pm\vec{k}}=7$, which is small enough to numerically
solve the master equation for the density matrix $\hat{\rho}:$
\begin{align}
\dot{\hat{\rho}}  &  =-i[H^{\prime(T)},\hat{\rho}]+\nonumber\\
&  \sum_{\vec{k}^{\prime}\in\{0,\vec{k},-\vec{k}\}}[\zeta_{mp,\vec{k}^{\prime
}}n_{th,\vec{k}^{\prime}}+\zeta_{mm,\vec{k}^{\prime}}n_{th,\vec{k}^{\prime}%
}^{\prime}]L_{\vec{k}^{\prime}}(\hat{\rho})+\nonumber\\
&  \sum_{\vec{k}^{\prime}\in\{0,\vec{k},-\vec{k}\}} \left[\frac{\zeta_{mp,\vec
{k}^{\prime}}}{2}+\frac{\zeta_{mm,\vec{k}^{\prime}}}{2} \right] L_{\vec{k}^{\prime}%
}^{\prime}(\hat{\rho}), \label{eq35}%
\end{align}
where $L_{\vec{k}^{\prime}}$ and $L_{\vec{k}^{\prime}}^{\prime}$ are the
Lindblad operators governing the dissipation in the Born-Markov approximation
\cite{Carmichael1999}%
\begin{align}
L_{\vec{k}^{\prime}}  &  =c_{\vec{k}^{\prime}}\hat{\rho}c_{\vec{k}^{\prime}%
}^{\dag}+c_{\vec{k}^{\prime}}^{\dag}\hat{\rho}c_{\vec{k}^{\prime}}-\hat{\rho
}c_{\vec{k}^{\prime}}c_{\vec{k}^{\prime}}^{\dag},\label{eq36}\\
L_{\vec{k}^{\prime}}^{\prime}  &  =2c_{\vec{k}^{\prime}}\hat{\rho}c_{\vec
{k}^{\prime}}^{\dag}-c_{\vec{k}^{\prime}}^{\dag}c_{\vec{k}^{\prime}}\hat{\rho
}-\hat{\rho}c_{\vec{k}^{\prime}}^{\dag}c_{\vec{k}^{\prime}}. \label{eq37}%
\end{align}
Eq. (\ref{eq35}) can be written in terms of a superoperator matrix
$\mathcal{L}$ as $\dot{\hat{\rho}}=\mathcal{L}\hat{\rho}$. The steady state of
the density matrix, $\rho_{ss}$, satisfies $\dot{\hat{\rho}}_{ss}%
=\mathcal{L}\hat{\rho}_{ss}=0$. Therefore, we search for the eigenvector with
zero eigenvalue of the superoperator matrix $\mathcal{L}$. This can be done in
two ways. First method: as the steady state equation $\mathcal{L}\hat{\rho}=0$
suggests, the process begins by reforming the matrix $\rho$ into a vector and
reforming $\mathcal{L}$ to conserve the equation. Then what remains is simply
getting the eigenvector corresponding to the smallest eigenvalue which is zero
by construction. This can be done using Lanczos loops. Second method: the
matrix $\mathcal{M}$ which gives $\mathcal{M}\rho=[Tr(\rho)=1,0,0,\dots]^{T}$,
is simply known. Using this matrix and $\mathcal{L}\rho=0$, we have
$(\mathcal{L}+\mathcal{M})\rho=\mathcal{M}\rho=[1,0,0,\dots]^{T}$. Now
$\rho_{ss}$ is reached by simply solving this linear system of equations. We
use the second method which is much faster, more scalable and more accurate.
We also use methods pertaining to sparse matrices in order to speed up and
maximize the computationally tractable Hilbert space.
%%%%%%%%%

\section{\label{app4}Entanglement measures}

%%%%

\subsection{\label{app4_1}Logarithmic negativity}

A bipartite system (\textquotedblleft Alice and Bob\textquotedblright) is
separable when $\hat{\rho}=\sum_{i}\eta_{i}\hat{\rho}_{i,1}\otimes\hat{\rho
}_{i,2}$, where $\eta_{i}$ is a coefficient, $\hat{\rho}$ is the total density
operator of the mixed state of system 1 and 2, $\hat{\rho}_{i,1(2)}$ is the
density operator of system 1 (2) for the pure state $i,$ and $\otimes$ denotes the
direct product. Here, for introduction, we focus on a pure state $\hat{\rho
}=\hat{\rho}_{1}\otimes\hat{\rho}_{2}$, since the treatment can be easily
extended to mixed states. Since $(\hat{\rho}_{1})^{T}=\hat{\rho}_{1}^{\ast}$
is a well behaved density matrix with positive eigenvalues, a negative
eigenvalue of $\hat{\rho}^{\prime}=(\hat{\rho}_{1})^{T}\otimes\hat{\rho}_{2}$
implies that the state would not be separable into its parts 1 and 2; in other
words the state would be entangled. This negative partial transpose (NPT)
criterium is generally a sufficient condition for entanglement
\cite{Peres1996,Horodecki1996,Braunstein2005}. Vidal and Werner
\cite{Vidal2002} introduced an entanglement measure based on NPT that
quantifies the degree with which $\hat{\rho}^{\prime}$ fails to be positive.
The trace norm $\left\Vert \rho^{\prime}\right\Vert _{1}=\mathrm{tr}\sqrt
{\rho^{\prime\dag}\rho^{\prime}}=1+2\left\vert \sum_{i}\mu_{i}\right\vert
=1+2\mathfrak{n}(\rho)$, where the sum is over all negative eigenvalues
$\mu_{i}<0$ of the partially transposed density matrix, and $\mathfrak{n}%
(\rho)$ is referred to as \textit{negativity}. The \textit{logarithmic
negativity} $E_{LN}=\text{log}_{2}||\rho^{\prime}||_{1}\geq E_{D}$ bounds
$E_{D},$ the rate at which entanglement can be distilled using local
operations and classical communications, the so-called LOCC
\cite{Bennett1996,Vidal1999,Vidal2000,Braunstein2005}. For example,
$E_{D}=0.4$ means that 10 copies of the state can in principle produce 4
perfect Einstein-Podolsky-Rosen (EPR) states \cite{Einstein1935}. EPR states
are non-local entangled states shared between two distinct particles (modes).
Examples are spin singlet states and two-mode squeezed states.

 When evaluating e.g. bipartite entanglement, a system of modes should be divided in several parts, each part consisting of one or more modes. It should be noted that the photon mode forms a hybridized mode (polariton) with the Kittel mode, and this polariton should be considered as one mode. Therefore, here we are dealing with essentially a maximally tripartite system, the Kittel mode-photon polariton and
the two modes of the instability driven $\pm\vec{k}$ pair. In this case, one should evaluate entanglement between
each part $\rho_{0p}$ (`$0p$' in the subscript indicates the part
corresponding to Kittel-photon polariton), $\rho_{-\vec{k}}$, and $\rho
_{\vec{k}}$ with the other two parts joined as one part, $\rho_{\vec{k}%
,-\vec{k}}=\hat{\rho}_{\vec{k}}\otimes\hat{\rho}_{-\vec{k}}$, $\rho
_{0p,\vec{k}}$, and $\rho_{0p,-\vec{k}}$. Therefore, there are only two
distinct configurations as shown in Fig. \ref{fig6}, with corresponding
logarithmic negativities $E_{LN,\pm\vec{k}\{0p,\mp\vec{k}\}}$ and
$E_{LN,0p\{\pm\vec{k}\}}$. From here on and in the main text, we drop `p' in
`0p' for simplicity.

The covariance matrix $\Lambda$ in Appendix \ref{app2} can be used to
calculate $E_{LN,\pm\vec{k}\{0p,\mp\vec{k}\}}$ and $E_{LN,0p\{\pm\vec{k}\}}$
\cite{Vidal2002}. For $E_{LN}$ calculated from covariance matrices, we add a
superscript `L', i.e. $E_{LN}^{L}$. Transposition corresponds to time
reversal, i.e. reversing the sign of momentum. The partially transposed
covariance matrix $\Lambda^{\prime}$ is obtained from $\Lambda$ by negating
the elements connecting momentum of the mode (modes) of one part to the
positions of the same part as well as to positions and momenta of the modes
of the other parts. For example, for evaluation of $E^{L}_{LN,\pm\vec{k}%
\{0p,\mp\vec{k}\}}$, $[\Lambda^{\prime}%
]_{1(3,4,5,6,7,1,3,4,5,6,7),2(2,2,2,2,2,8,8,8,8,8,8)}=-[\Lambda
]_{1(3,4,5,6,7,1,3,4,5,6,7),2(2,2,2,2,2,8,8,8,8,8,8)}$, where $[\Lambda^{(\prime)}]_{1(3,4,\cdots),2(2,2,\cdots)}$ indicates $[\Lambda^{(\prime)}]_{1,2}$, $[\Lambda^{(\prime)}]_{3,2}$, $[\Lambda^{(\prime)}]_{4,2}$ $\cdots$, respectively, and
$\Lambda^{\prime T}=\Lambda^{\prime}$. The logarithmic negativities for
Gaussian distributed states \cite{Vidal2002} are
\begin{equation}
E_{LN,i}^{L}=\sum_{i=1}^{8}\mathcal{Y}\left(  \mathfrak{d}_{i}\right)  ,
\end{equation}
where $i$ is either $\pm\vec{k}\{0,\mp\vec{k}\}$ or $0\{\pm\vec{k}\}$,
$\mathcal{Y}(\mathfrak{d}_{i})=-\mathrm{log}_{2}(2\mathfrak{d}_{i})$ if
$2\mathfrak{d}_{i}<1$ and zero otherwise, $\mathfrak{d}_{i}$ is an eigenvalue
of $\sigma^{-1}\Lambda^{\prime}$, and $\sigma$ is a $8\times8$ block-diagonal
matrix with blocks formed by the Pauli matrix $\sigma_{x}$. This evaluation
fails for states that are far from being Gaussian.

We also evaluate $E_{LN}$ from the density matrix of the steady states
$\rho_{ss}$ achieved from solving the quantum master equation corresponding to
the scaled system as described in Appendix \ref{app3}. This is a
straightforward calculation, as one only requires to transpose the $\rho_{ss}$
components corresponding to the Hilbert space of one of the parts for each
bipartite configuration. For $E_{LN}$ calculated from density matrices, we add
a superscript `q', i.e. $E_{LN}^{q}$. $\rho_{ss}$ is a matrix with entries
corresponding to $\vert i,j,k\rangle\langle i^{\prime},j^{\prime},k^{\prime
}\vert$, $i$ ($i^{\prime}$), $j$ ($j^{\prime}$), and $k$ ($k^{\prime}$) refer
to the $i$'th ($i^{\prime}$'th), $j$'th ($j^{\prime}$'th), and $k$'th
($k^{\prime}$'th) Fock (number, level) state of the Kittel mode, $\vec{k}$
mode, and $-\vec{k}$ mode, respectively. Now, for obtaining e.g.
$E^{q}_{LN,0\{\pm\vec{k}\}}$, we form partial transposed density matrix
$\rho_{ss}^{PT}$ with entries $\vert i^{\prime},j,k\rangle\langle i,j^{\prime
},k^{\prime}\vert$ equivalent to entries $\vert i,j,k\rangle\langle i^{\prime
},j^{\prime},k^{\prime}\vert$ of $\rho_{ss}$. Now diagonalize $\rho_{ss}^{PT}%
$, and obtain its negative eigenvalues which lead directly to $E^{q}%
_{LN,0\{\pm\vec{k}\}}$. Similar partial transposition leads to $E_{LN,\pm
\vec{k}\{0,\mp\vec{k}\}}^{q}$.
%%%%

\subsection{\label{app4_2}Entanglement of formation}

The logarithmic negativity gives an upper bound for distillable entanglement
$E_{D}$ of a bipartite state. Even if $E_{D}=0$ for a general mixed state, it
does not mean that there is no entanglement in that state. This concept
manifests itself in the reverse process, i.e. the required number of
completely entangled particles for building a certain bipartite state, which
has a different measure named entanglement of formation, $E_{F}$
\cite{Bennett1996_1,Wootters2001}. This quantity is the same as $E_{D}$ for
pure states \cite{Bennett1996}. In this case, both are equivalent to the von Neumann
entropy $E_{vN}=\text{Tr}\rho_{1(2)}\log_{2}{\rho_{1(2)}}$ where $\rho
_{1(2)}=\text{Tr}_{2(1)}{\rho}$ (tracing over the part 2 (1) of the bipartite
state), and $\rho$ is the total density matrix of the bipartite system and
$\rho_{1(2)}$ is the density matrix of part 1 (2). However, when we deal with
mixed states, there is no closed formula for $E_{F}$. For a bipartite mixed
state, $E_{F}$ is the minimum $E_{vN}$ among different realizations of a mixed
state using pure states $\rho=\sum_{i}p_{i}\vert \Upsilon_{i}\rangle
\langle \Upsilon_{i}\vert$, where $\Upsilon_{i}$ is a pure state. There are
infinite ways of assigning an ensemble of $\left\{  p_{i},\Upsilon
_{i}\right\}  $ for a mixed state. Even though there are analytical
expressions for mixed state of two qubits \cite{Wootters2001} as well as
approximate closed form solution and bounds for two-mode Gaussian states
\cite{Adesso2005}, evaluating $E_{F}$ for a general system can only be done
using numerical methods. We utilize the algorithm given in Ref.
\cite{Zyczkowski1999} for fnding $E_{F}$. Here, we give a summary of the
algorithm we used: \newline(1) Perform a singular value decomposition on the
mixed state $\rho=U_{\rho}\times S_{\rho}\times V_{\rho}$, where $S_{\rho}$ is
a diagonal matrix containing all the singular values in descending order. As
the numerical method in Ref. \cite{Zyczkowski1999} becomes less effective at
large dimensions of $\rho$, we should perform a cutoff much similar to the one
performed in density matrix renormalization group calculation method
\cite{White1992,*White1993,*Schollwock2011}. We keep $N_{co}$ largest values
of $S_{\rho}$ and discard the rest, and also discard the corresponding columns
of $U_{\rho}$ and corresponsing rows of $V_{\rho}$, to reach $S_{\rho}%
^{\prime}$, $U_{\rho}^{\prime}$ and $V_{\rho}^{\prime}$, respectively.
Therefore, we have reached a decomposition of the mixed state $\rho$ by pure
states in a space of smaller dimension, $\rho=\sum_{i}^{N_{co}}p_{i}%
\vert \Upsilon_{i}\rangle\langle \Upsilon_{i}\vert$, wehere $p_{i}$ are entries
of $S_{\rho}^{\prime}$ and $\vert \Upsilon_{i}\rangle$ are rows of $V_{\rho
}^{\prime}$. \newline(2) Form a random unitary matrix $\mathcal{U}$ of
dimension $N_{co}\times N_{co}$, with the method detailed in Ref.
\cite{Pozniakyk1998}. Subsequently, form the density matrix $\rho^{\prime
}=\sum_{i}^{N_{co}}\vert \Upsilon^{\prime}_{i}\rangle\langle \Upsilon^{\prime
}_{i}\vert$, where $\vert \Upsilon^{\prime}_{i}\rangle=\sum_{j}^{N_{co}%
}\mathcal{U}_{ji}\vert \Upsilon_{i}\rangle$. \newline(3) Evaluate $E_{vN}$ for
$\rho^{\prime}$, and memorize its value as $E_{vN,0}$. \newline(4) Form a
random hermitian matrix $\mathcal{R}$ with each of the matrix entries of row
(column) i (j) having Gaussian distribution with variance $(1\pm\delta
_{ij})/N_{co}$ for real and imaginary parts, respectively. \newline(5) Form
the unitary matrix $\mathcal{U}^{\prime}=\mathcal{U}\exp\left[  i\chi
\mathcal{R}\right]  $. If it is the first time for this step, set $\chi
=\chi_{0}$, if not, scale down $\chi$ by $\Xi<1$. \newline(6) Form
$\rho^{\prime}$ corresponding to $\mathcal{U}^{\prime}$ similar to what was
described in step (2). Evaluate the corresponding $E_{vN}$, and memorize its
value as $E_{vN,1}$. \newline(7) If $E_{vN,1}<E_{vN,0}$, accept the move, i.e.
set $E_{vN,0}=E_{vN,1}$ and $\mathcal{U}=\mathcal{U}^{\prime}$. \newline(8)
Repeat steps (5)-(7), $N_{\chi}$ times. Memorize all the accepted values of
$E_{vN,1}$. \newline(9) Repeat steps (4)-(8), $N_{\mathcal{R}}$ times.
\newline(10) Repeat steps (2)-(9), $N_{\mathcal{U}}$ times. \newline(11) The
minimum among all the memorized values of $E_{vN,1}$ is the entanglement of
formation $E_{F}$.

\begin{figure}[!]
\includegraphics[width=0.5\textwidth]{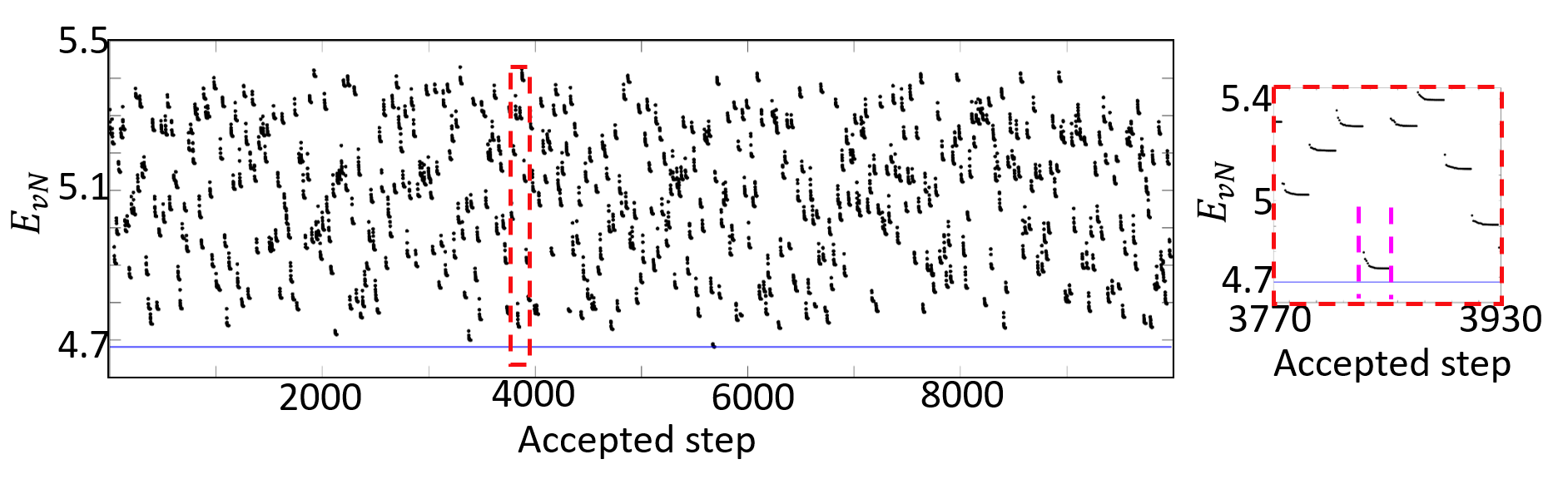}\caption{The von Neumann
entropy of each accepted step ($E_{vN,1}$) in the process for calculation of
entanglement of formation $E_{F}$. The right panel is a zoom in of the area
within the red-dashed rectangle in the left panel. The points in between the
purple dashed lines belong to minimizing process for a certain initial random
unitary matrix.}%
\label{fig_SI_EF}%
\end{figure}

We only calculate $E_{F,0\{\pm\vec{k}\}}$, and for that we set $N_{co}=8$,
$N_{\chi}=20$, $N_{\mathcal{R}}=20$, $N_{\mathcal{U}}=500$, $\chi_{0}=0.3$,
$\Xi=2/3$. In Fig. \ref{fig_SI_EF}, we show $E_{vN}$ for accepted moves, i.e.
memorized $E_{vN,1}$, in the case of injection locking pumping amplitude
${B}_{l}=10^{12}$ (see Fig. \ref{fig8}). The large spread of $E_{vN}$ for
possible decompositions of the mixed state into pure states shows how delicate
finding $E_{F}$ for a general mixed state is. In the right panel of Fig.
\ref{fig_SI_EF}, a zoomed in section of the left panel is shown, and the data
points surrounded by purple dashed-lines correspond to memorized $E_{vN,1}$
values in $N_{\mathcal{R}}$ iterations of steps (4)-(8) for a certain initial
random unitary matrix generated in step (2). The blue line in Fig.
\ref{fig_SI_EF} indicates the minimum among all values of accepted $E_{vN,1}$,
i.e. the entanglement of formation $E_{F}$.

\end{document}